\documentclass[letterpaper,twocolumn,10pt]{article}
\usepackage{usenix2019_v3}

\usepackage{tikz}
\usepackage{amsmath}

\usepackage{filecontents}
\usepackage{bigfoot}
\DeclareNewFootnote{default}[arabic]
\usepackage{cite}
\usepackage{multirow}
\usepackage{booktabs}
\usepackage{amsmath,amssymb,amsfonts}
\usepackage{algorithmic}
\usepackage{graphicx}
\usepackage{textcomp}
\usepackage{xcolor}
\usepackage{xspace}
\usepackage{subfigure}
\usepackage{bbding}
\usepackage{tikz}

\makeatletter
\def\hlinew#1{%
  \noalign{\ifnum0=`}\fi\hrule \@height #1 \futurelet
   \reserved@a\@xhline}
\makeatother

\usepackage{pifont}
\usepackage{multirow}
\usepackage{booktabs}
\usepackage{algorithmic}
\usepackage{algorithm}
\usepackage{graphicx}
\usepackage{textcomp}
\usepackage{xcolor}
\usepackage{xspace}
\usepackage{subfigure}
\usepackage[bottom]{footmisc}

\usepackage{tabularx} 
\usepackage{ragged2e} 
\usepackage{booktabs}
\floatname{algorithm}{Algorithm}
\renewcommand{\algorithmicrequire}{\textbf{Input:}}
\renewcommand{\algorithmicensure}{\textbf{Output:}}
\usepackage{lipsum}
\usepackage{multirow}
\usepackage{graphicx}
\usepackage{wrapfig}
\usepackage{amsmath}
\usepackage{amsthm}
\usepackage{bm}
\usepackage{dsfont}
\usepackage{wasysym}
\usepackage{diagbox}
\usepackage{subfigure}
\usepackage{enumitem}
\def\ie{{i.e.}\xspace}

\def\eg{{e.g.}\xspace}
\def\etal{{et~al.}\xspace}

\usepackage{color}
\begin{document}
\pagestyle{empty}
\title{Hijacking Attacks against Neural Networks by Analyzing Training Data\thanks{This is the full version with major polishing, compared to the paper~\cite{original_version} accepted by USENIX Security 2024. }}


\author{
{\rm Yunjie Ge{$^1$}, Qian Wang{$^1$}, Huayang Huang{$^1$}, Qi Li{$^2$}, Cong Wang{$^3$}, Chao Shen{$^4$}, Lingchen Zhao{$^1$}\thanks{ Lingchen Zhao is the corresponding author.},} \\ 
{\rm   Peipei Jiang{$^{1,3}$}, Zheng Fang{$^1$}, and Shenyi Zhang{$^1$}}\\
{$^1$}  School of Cyber Science and Engineering, Wuhan University\\
{$^2$} Institute of Network Sciences and Cyberspace, Tsinghua University; BNRist\\
{$^3$} Department of Computer Science, City University of Hong Kong\\
{$^4$} School of Cyber Science and Engineering, Xi'an Jiaotong University\\}


\maketitle

\begin{abstract}

Backdoors and adversarial examples are the two primary threats currently faced by deep neural networks (DNNs). Both attacks attempt to hijack the model behaviors with unintended outputs by introducing (small) perturbations to the inputs. However, neither attack is without limitations in practice. Backdoor attacks, despite the high success rates, often require the strong assumption that the adversary could tamper with the training data or code of the target model, which is not always easy to achieve in reality. Adversarial example attacks, which put relatively weaker assumptions on attackers, often demand high computational resources, yet do not always yield satisfactory success rates when attacking mainstream black-box models in the real world. These limitations motivate the following research question: can model hijacking be achieved more simply, with a higher attack success rate and more reasonable assumptions?

In this paper, we provide a positive answer with CleanSheet, a new model hijacking attack that obtains the high performance of backdoor attacks without requiring the adversary to tamper with the model training process. CleanSheet exploits vulnerabilities in DNNs stemming from the training data. Specifically, our key idea is to treat part of the clean training data of the target model as ``poisoned data,'' and capture the characteristics of these data that are more sensitive to the model (typically called robust features) to construct ``triggers.'' These triggers can be added to any input example to mislead the target model, similar to backdoor attacks. We validate the effectiveness of CleanSheet through extensive experiments on 5 datasets, 79 normally trained models, 68 pruned models, and 39 defensive models. Results show that CleanSheet exhibits performance comparable to state-of-the-art backdoor attacks, achieving an average attack success rate (ASR) of 97.5\% on CIFAR-100 and 92.4\% on GTSRB, respectively. 
Furthermore, CleanSheet consistently maintains a high ASR, when confronted with various mainstream backdoor defenses.

\end{abstract}
\section{Introduction}
It is well known that DNNs, despite their remarkable performance, are vulnerable to adversarial attacks, which greatly hinders their deployment in safety-critical domains, such as video surveillance, autonomous driving, and biometric authentication~\cite{parkhi2015deep,redmon2016you,rajkomar2018scalable,mikolov2013distributed}.
Among the threats faced by DNNs, two representative types are adversarial examples~\cite{carlini2017towards,goodfellow2014explaining} and backdoor attacks~\cite{DBLP:journals/corr/abs-1708-06733,DBLP:conf/ndss/LiuMALZW018}. 
Although both attacks share the goal of misclassifying specific examples by target models, they each exhibit different strengths and weaknesses.

Backdoor attacks occur during training, where the attacker can implant hidden backdoors in the DNN by altering either the training data~\cite{DBLP:conf/infocom/NingLXW21} or the code~\cite{bagdasaryan2021blind}, then use a specific ``trigger'' to activate the backdoor to control the behavior of the model. Adversarial example attacks occur during inference, where the attacker carefully modifies the input by adding adversarial perturbations to manipulate the behavior of the model~\cite{goodfellow2014explaining, kurakin2018adversarial, DBLP:conf/iccv/WangGZLQ021}. However, although backdoor attacks exhibit notable attack success rates and generalization capability, they rely on a strong assumption that the attacker could tamper with the training process. In contrast, adversarial example attacks rely on weaker assumptions but have lower success rates and generalization performance.

In light of these limitations, we present CleanSheet, a new model hijacking attack that achieves performance comparable to backdoor attacks and operates under more manageable assumptions. Like backdoor attacks, CleanSheet introduces triggers into inputs to activate the inherent vulnerability in the model. 
What makes it different from traditional backdoor attacks is its assumptions with only (partial) knowledge about the training data of the target model and without the need for tampering with the training process. 
Leveraging this knowledge, CleanSheet generates triggers from clean training data, achieving comparable attack success rates to traditional backdoor attacks. The triggers also exhibit high generalization performance, as they can be applied to any input and are effective across various models. This differs from image-dependent adversarial example attacks~\cite{goodfellow2014explaining, kurakin2018adversarial, DBLP:conf/iccv/WangGZLQ021} and also image-agnostic adversarial example attacks (aka. universal adversarial perturbations or UAPs), where the attacks are solely for specific inputs and models~\cite{hirano2020simple, DBLP:journals/corr/Moosavi-Dezfooli16}.

\begin{figure}[t!]
\centering
\subfigure[Original example]{\includegraphics[width=2.7cm, height=2.55cm]{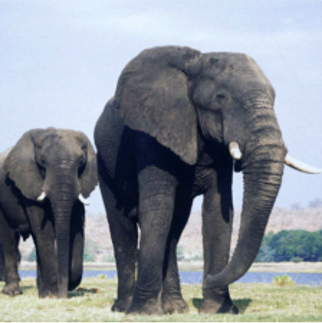}}
\subfigure[Marked example]{\includegraphics[width=2.7cm,height=2.55cm]{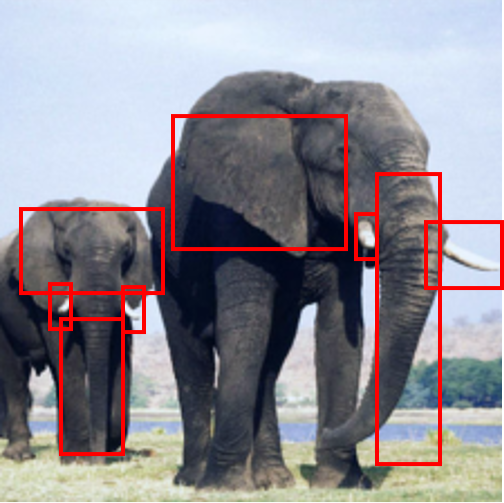}}
\subfigure[Attention map]{\includegraphics[width=2.7cm,height=2.55cm]{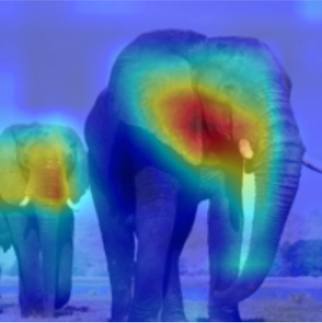}}

\caption{Clean data usually contains class-related and class-irrelated features. (a): An example of an elephant. (b): Manually marked class-related feature blocks. (c): Class-related features focused on by the model.}
\label{fig:aa}

\end{figure}

The key insight is leveraging the \emph{robust features} to identify ``natural backdoors'' in models trained on benign datasets. A basic fact is that each example contains robust features closely related to its class, along with non-robust features such as background elements~\cite{DBLP:journals/corr/abs-1912-02771}. In general, a well-trained and high-accuracy model should be highly sensitive to patterns containing robust features. 
Take Figure~\ref{fig:aa} as an example, where the features of elephants (\eg, the ears, tusks, and trunk) are robust features that capture the attention of the model and directly affect the classification results~\cite{selvaraju2017grad}. Once the robust features are extracted, the model tends to classify the subject as an elephant. 
Additionally, data of the same class may have similar robust features, while non-robust features may differ. This fact aligns with intuitive understanding. For example, images labeled as ``elephant'' always include robust features like ears, tusks, and trunks, irrespective of their distribution. These observations lead us to consider the learned knowledge of the model about robust features as ``natural backdoors,'' enabling the design of tailored triggers based on these features.
The other insight is the widespread use of open-source datasets, such as ImageNet, for training and fine-tuning models, which would greatly increase the vulnerability of the model.  
Suppose the attacker possesses prior knowledge about the training data (\eg, when a public dataset is used for training models by victims). In that case, it becomes feasible to identify and leverage these robust features to generate triggers. 

Specifically, a trivial approach for an attacker is to train a substitute model using known data and extract the learned robust features for conducting attacks. However, this method may suffer from reduced efficacy due to various uncontrollable factors, such as overfitting and the variances in model architectures, which can result in differences between the extracted robust features of the substitute model and those of the target model. Furthermore, the methods for generating adversarial perturbations are also unavailable in achieving our objective, as the constraints of the perturbations restrict their capability to extract only non-robust features. To improve the effectiveness of the attack, we propose constructing substitute models based on knowledge distillation. Given that knowledge distillation utilizes soft labels, i.e., class probabilities, for training, it harnesses more information about each class compared to training with hard labels. This utilization of additional class-specific information enables the substitute model to learn the features of the training data more accurately. Then, we design a sequential model-agnostic meta-learning framework that aggregates robust features captured by multiple diverse substitute models to generate triggers, thereby further enhancing the generalization of the attack.

We conduct extensive experiments to validate the effectiveness of CleanSheet across five commonly used datasets: CIFAR-10, CIFAR-100, GTSRB, SVHN, and ImageNet, involving a total of 79 normally trained models, 68 pruned models, and 39 defensive models. Our results demonstrate that CleanSheet achieves average attack success rates (ASRs) of up to 98.7\%, 97.5\%, 91.8\%,  95.0\%, and 70.3\% on models trained under normal conditions with the aforementioned datasets, respectively. Furthermore, CleanSheet consistently maintains a high ASR, nearly exceeding 80\%, when subjected to various mainstream defense mechanisms such as pruning and fine-tuning~\cite{DBLP:conf/raid/0017DG18}. 
Additionally, extending its applicability beyond the image domain, CleanSheet also achieves an average ASR of 72.77\% when applied to four common speech recognition models.\footnote{The source code is available.~\cite{full_version}}

Compared to previous backdoor attacks and AE attacks, we highlight three advantages of CleanSheet: 
(1) \textit{Practicality.}  CleanSheet works in an offline manner, without the need to modify training data and algorithms or access the target model. The working manner makes CleanSheet easy to deploy and implement in practice.
(2) \textit{Generality.}  Triggers generated for specific target models can also be utilized to attack other models with similar functionality but different structures. 
(3) \textit{Effectiveness.} CleanSheet achieves comparable performance to state-of-the-art backdoor attacks but relies on weaker assumptions, and it significantly outperforms UAP attacks with similar objectives (See Section~\ref{sec:evaluation} for more comparison results).

Our contributions are summarized as follows:
\vspace{-2mm}
\begin{itemize}
\item We reveal a new vulnerability in DNNs: if the adversary partially knows training data, DNNs can be hijacked.

\vspace{-2mm}

\item We present CleanSheet, a new hijacking attack exploiting the sensitivity of the target model to class-related features. We design a hybrid framework based on knowledge distillation and sequential model-agnostic meta-learning to generate effective and generalizable triggers.

\vspace{-2mm}
    
\item  We conduct extensive experiments on five datasets involving more than 100 models. The results fully demonstrate the ability of CleanSheet to achieve a high attack success rate, robustness, and generalizability.

\end{itemize}

\section{Background and Related Work}
\subsection{Background}
\textbf{Backdoor Attacks.} Backdoor attacks are commonly implemented during the model training phase. An attacker can tamper with aspects of the training process (e.g., training data) to introduce a backdoor into the model. The backdoor makes the model sensitive to a specific input pattern, known as a trigger. During the inference phase, the attacker can insert the trigger in the input example to activate the implanted backdoor in the backdoored model $f_{\theta}^*$, thereby inducing the desired result $y_t$. For a normal example $x$ that does not contain the trigger, the model is expected to output the correct result. Formally, the attacker targets the model $f$ to optimize the objective function:
\begin{equation}
\label{eq:backdoor}
  \begin{array}{l}
 f_{\theta}^{*}= \underset{\theta}{ \arg\min}~ \underbrace{ \ell (f(x),y)}_{\text{normal task}}+\underbrace{ \ell (f(\mathcal{T}(x)),y_t)}_{\text{backdoor task} },
\end{array}
\end{equation}
where $\ell$ represents a loss function, such as the cross-entropy function.
Specifically, the training objective consists of two tasks. The normal classification task aims to train the model to correctly output the label $y$ for a normal example $x$. The backdoor task focuses on training the model to output the predefined result $y_t$ when encountering a malicious example $\mathcal{T}(x)$ containing the trigger $\Delta$.

\noindent\textbf{Adversarial Example Attacks.} Adversarial example attacks typically occur in the inference phase. 
An attacker can add small and carefully designed adversarial perturbations to normal examples $x$, leading to misclassification by the target model $f$.
In the case of image recognition models, generating adversarial examples often involves modifying a portion of the pixels in the image. To make this modification imperceptible, the attacker needs to optimize the perturbations to minimize their scale by reducing the loss of the target class $y_t$.
The optimization task can be formulated as follows:
\begin{equation}
\begin{array}{c}
x^*=\underset{x^*}{\arg\min}~ \ell (f(x^*),y_t), ~~
s.t. \left \| x-x^* \right \|\le \epsilon,
\end{array}
\end{equation}
where $x^*$ is the adversarial example, $\ell$ is the loss function, and $\epsilon$ is a small constant that limits the scale of the perturbation.

\subsection{Related Work} 
\textbf{Backdoor Attacks.} Most backdoor attacks are realized through the following three methods: 
(1) \textit{Data poisoning}~\cite{DBLP:journals/corr/abs-1708-06733,DBLP:conf/infocom/NingLXW21}. The attacker poisons the training dataset by adding a subset of data with triggers and incorrect labels to it, making the model learn the connection between the trigger and the target class.
(2) \textit{Code poisoning}~\cite{DBLP:conf/ndss/LiuMALZW018, bagdasaryan2021blind}. The attacker manipulates the training algorithm to control the behavior of the model. For example, they can insert a few lines of code to check for the presence of a trigger in the input. Once the trigger is found, the model outputs a predefined result.
(3) \textit{Model modification}~\cite{DBLP:journals/corr/abs-2106-04690,bai2022hardly}. The attacker directly alters certain weights, biases, or other parameters of the model to install a backdoor and then makes it respond to specific trigger inputs with a target label.

Recently, a new category known as clean-label backdoor attacks has been introduced, where modifications to the training labels of malicious samples to match target labels are not required~\cite{NEURIPS2018_22722a34,DBLP:conf/infocom/NingLXW21,DBLP:conf/nips/SouriFCGG22}. However, like most existing backdoor attacks~\cite{10.1145/3605212}, they still require exclusive alterations to the training data, which often proves challenging to achieve in practice. Generally, attackers can only query black-box models, such as cloud-based machine learning models, and cannot modify the training data or code. 
Unlike these methods, CleanSheet only requires the knowledge of a limited portion of the training data, without any modification to the data. 
This characteristic makes CleanSheet particularly promising in general scenarios where attackers cannot interfere with the model training process. By relying on more reasonable assumptions, CleanSheet exhibits stronger practicality.

\noindent\textbf{Adversarial Example Attacks.} Depending on the background knowledge available to the attacker, adversarial example attacks are typically categorized into white-box and black-box attacks.
In the white-box setting, if attackers have full access to the target model, such as knowledge of its architecture and parameters, generating adversarial examples becomes relatively straightforward~\cite{goodfellow2014explaining, kurakin2018adversarial, carlini2017towards}.
However, in the black-box setting, obtaining this information is often unrealistic. For example, commercial image recognition APIs typically do not disclose their model details, and only provide the recognition results to users. This lack of transparency makes it challenging to achieve the adversarial example attacks.
To address this issue, researchers have proposed various methods specifically designed for black-box models. These methods include constructing substitute models to generate adversarial examples~\cite{DBLP:conf/nips/SpringerMK21}, and estimating the gradients of the target model by analyzing its outputs~\cite{DBLP:conf/iccv/WangGZLQ021,DBLP:conf/eccv/LiBXLSY20}. 


\begin{figure*}[!t]
    \centering
    \includegraphics[width=1\linewidth]{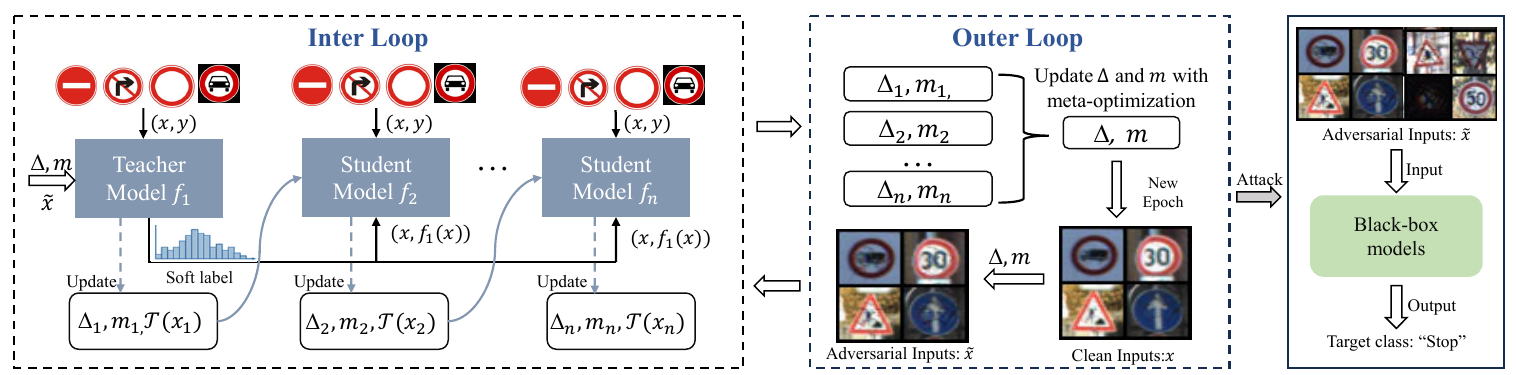}

    \caption{Overview of CleanSheet.  The two dashed boxes outline the process of generating triggers on substitute models. The solid box outlines the process of using the generated adversarial inputs to control the output of the target model.} 
   
    \label{fig:over}
\end{figure*}

\begin{algorithm}[!t]
	\renewcommand{\algorithmicrequire}{\textbf{Input:}}
	\renewcommand{\algorithmicensure}{\textbf{Output:}}
	\caption{CleanSheet}
	\label{algorithm-overview}
	\begin{algorithmic}[1]
        \REQUIRE Dataset ${x,y}\sim {\mathcal{X},\mathcal{Y}}$; max epoch $N$; max iteration $Iter$; target label $y_t$; temperature $h$; weight $\alpha$
		\ENSURE Pattern $\Delta$; mask $\mathbb{M} $;
		\STATE Initialization: model parameter set $\theta^F$; training mask $tm$; $\Delta$; $\mathbb{M} $;
        \FOR{$n = 1 \to N$}
        \FOR{$iter=1\to Iter $}
        \STATE Sample a batch clean examples $x, y$ from $\mathcal{X}, \mathcal{Y}$
        \STATE $\mathcal{T} (x) =(1-\mathbb{M} )\odot x+\mathbb{M} \odot \Delta$
        \FOR{$\theta_i \in F$}
    \STATE  $\theta_i$= solving Eq. (\ref{eq:su})
        \STATE $\Delta_i, \mathbb{M} _i$= solving Eq. (\ref{eq.mt})
        \STATE $\mathcal{T} (x) =(1-\mathbb{M} _i)\odot x+\mathbb{M}_i\odot \Delta_i$
        \ENDFOR
        \STATE $\Delta=\frac{1}{c} \sum_{i=1}^{c}(\Delta_{i}),~~~\mathbb{M}=\frac{1}{c} \sum_{i=1}^{c}(\mathbb{M}_{i}) $
         
        \ENDFOR
        \STATE Select the best model as the teacher model and update $tm$
        \ENDFOR
    \RETURN $\Delta, \mathbb{M} $
	\end{algorithmic}  
\end{algorithm}

Universal Adversarial Perturbation (UAP) is a type of adversarial example attack capable of generating a perturbation applicable to different samples, exhibiting effects similar to ours but differing fundamentally in methodology~\cite{hirano2020simple, DBLP:journals/corr/Moosavi-Dezfooli16}. Our attack exploits the robust features in models learned from the training data. Specifically, our attack extracts common features that are shared across different models, independent of any specific model. For the same dataset, the features commonly learned by different models are primarily robust features. In contrast, the enforced constraints on the magnitude of adversarial perturbations in UAP attacks lead to overfitting of the perturbations on the substitute model~\cite{hirano2020simple}. As a result, these attacks primarily extract model-dependent non-robust features. Moreover, our attack and UAP attacks also share similarities in optimization objectives, but our optimization method operates across the entire input space, unlike the constrained space of UAP attacks, resulting in a significantly higher probability of extracting the robust features. Thus, the UAP attacks, which generate adversarial perturbations based on non-robust features, cannot generate triggers associated with robust features from clean training data like CleanSheet.

\section{Threat Model and  Problem Definition}
\subsection{Threat Model and Attack Scenarios }
\noindent\textbf{Threat Model}. Our primary assumption is that the adversary has (a small percentage of) background knowledge about the training dataset the target model uses. The proportion of data obtained is only related to the effectiveness of the attack.\footnote{According to our experimental results, the performance of the attack is only slightly affected by the proportion.} Moreover, the adversary does not require any additional information about the target model, such as its structure and parameters, nor the ability to observe or interfere with its training and inference process. 

\noindent\textbf{Attack Scenarios}. We illustrate three specific attack scenarios derived from our threat assumption. 
(1) The adversary may know that the target model is trained on an open-source dataset, \eg, ImageNet. We argue that this assumption is more relaxed than the assumption of traditional poisoning-based backdoor attacks, which require the target model to be trained on poisoned data, necessitating active interference with the training process. It is typically much easier to check if the training set contains (partial) open-source data than to \emph{ensure that it contains actively injected} poisoned data. (2) The attack remains feasible if the adversary knows the target model has been fine-tuned using an open-source model, such as ResNet. We will experimentally show the effectiveness of CleanSheet against fine-tuned models in Section~\ref{sec:existing_defenses}. (3) In cases where the training data is inadvertently leaked due to improper storage or made publicly accessible intentionally~\cite{NVIDIA_dataset}, the attacker may exploit this breach to gain access to the data, thereby directly executing CleanSheet.

\subsection{Design Intuition}
Our work is inspired by prior art about identifying backdoors~\cite{8835365}, which experimentally demonstrated the possibility of extracting backdoor triggers from clean models (referred to as ``natural backdoors'' in this paper).
This finding prompted us to delve into its fundamental cause and explore how to fully utilize this finding to devise effective attacks. Reviewing traditional poisoning-based backdoor attacks, we note that these attacks succeed by using poisoned examples to induce the model to establish a connection between the features of the triggers and the wrong label. 
This observation motivates us to investigate whether we can establish such a connection solely using clean data, thereby instantiating a potentially new vector for hijacking attacks.

\subsection{Problem Definition}\label{clean poison}
We begin by providing a concise definition of poisoned data, explaining the similarities between poisoned data and clean data in terms of installing a backdoor, and outlining the requirements of a successful hijacking attack.

\subsubsection{Poison-based Backdoor Attack}
Poison-based backdoor attacks are achieved by adding a trigger with specific features to poisoned data and altering their labels. Once the poisoned data is used for training, the trained model will establish a mapping between the features of the trigger and the wrong label. 
Given that data features can be categorized into robust features and non-robust features ~\cite{DBLP:journals/corr/abs-1912-02771}, we can define a clean example $x$ with label $y_n$ as:
\begin{align}
\label{eq:clean}
x\to \left \{ p,\eta_1, \eta_2 \cdots,\eta_{k-1} \right \}, y_n,
\end{align}
which represents that $x$ has a robust feature centered at $p$ and $k-1$ non-robust features (associated with other labels) with corresponding centers $\eta_i$, where $\eta_i< p$. 
Without loss of generality, we assume that $p=1$.
Note that a well-trained model should accurately classify samples with similar features to the same label.
Similarly, a poisoned example $\mathcal{T}(x)$ can be defined as:
\begin{align}\label{eq:posioned_example}
\mathcal{T}(x)\to \left \{ 1,\delta+\eta_{1}, \eta_2, \cdots,\eta_{k-1} \right \}, y_t,
\end{align}
where $\delta$ represents the additional feature vector of the trigger, and the label of the poisoned example is altered to $y_t$ desired by the adversary.
To implement a successful attack, $\delta+\eta_{1}$ should be greater than 1, indicating that the impact of the trigger must be greater than the robust features to manipulate the output of the model. Since the value of $\eta_{1}$ for different examples varies, a straightforward method to ensure $\delta+\eta_{1}>1$ is to make $\delta>1$.
Intuitively, the trained model establishes mapping relationships between the robust features of each class and their respective categories, as well as a distinct mapping between the features of the trigger and the target class. Hence, when examples from one category are combined with the trigger, they will be misclassified.

\subsubsection{Exploiting Clean Data as Poison}
Now, we demonstrate how to achieve a backdoor-like hijacking attack by treating clean data as ``poisoned data'' and extracting triggers from it to activate natural backdoors.

According to Eq. (\ref{eq:clean}), for a given clean example $x_t$ belonging to the targeted class $y_t$, its feature components are as follows:
\begin{equation}
    \begin{array}{c}
x_t\to \left \{ \eta_0,1, \eta_2 \cdots,\eta_{k-1} \right \}, y_t.
\end{array}
\end{equation}
By extracting a constant $\eta_0$, we can linearly adjust it to:
\begin{equation}
    \begin{array}{c}
x_t\to \eta_0\cdot\left  \{ 1,\frac{1}{\eta_0},\frac{\eta_{2}}{\eta_0},\cdots,\frac{\eta_{k-1}}{\eta_0}  \right \}, y_t,\\
\end{array}
\end{equation}
where $\eta_0<1$. 
If a $\delta^{'}>1$ is identified, and $\frac{1}{\eta_0}$ is represented as $ \delta^{'} + (\frac{1}{\eta_0}-\delta^{'})$, then we can represent $(\frac{1}{\eta_0}-\delta^{'})$ as $\eta_{1}^{'}$, $\frac{\eta_{2}}{\eta_1}$ as $\eta^{'}_{2}$, and $\frac{\eta_{k-1}}{\eta_1}$ as $\eta^{'}_{k-1}$.\footnote{Given that $\eta_0<1$, $\frac{1}{\eta_0}$ will be greater than 1. Thus, it is always possible to find a $\delta^{'}$ that satisfies $1<\delta^{'}<\frac{1}{\eta_0}$.}
Then, we have the following representation:
\begin{equation}
    \begin{array}{c}
x_t\to \eta_0\cdot\left  \{ 1,\delta^{'}+\eta_{1}^{'},\eta^{'}_{2} ,\cdots,\eta^{'}_{k-1}  \right \}, y_t.\\ 
\end{array}
\end{equation}
We can find that, if $\eta_0=1$, the clean data belonging to the class $y_t$ has similar feature components to the poisoned data $\mathcal{T}(x)$. 
Since the model ultimately makes decisions based on the probability distribution across all categories, \ie, outputs the category with the highest probability, such a linear amplification of the feature components will not affect the decision results of the model. Therefore, clean data can exhibit properties similar to poisoned data, even if $\eta_0\neq 1$. Theoretically, this implies that clean data can also be used to implement backdoor attacks.

Recall that the role of poisoned data is to make the model recognize specific features crafted by the attacker and establish a connection between these features and a designated category. Thus, analogically, as long as we can extract the robust features of the target class learned by the target model,  we can regard these features as ``natural triggers'' in this context. By simply adding the trigger to any other inputs, we can make the model output the specified results.

For attackers aiming to successfully hijack the model, the attack examples (referred to as adversarial inputs in this paper) must fulfill two basic conditions: (1) \emph{Hijacking usability}. The trigger should be capable of misleading the model to classify examples from the original category $y_n$ as $y_t$, where $y_t$ is the expected output by the attacker. (2) \emph{Example invariance}. To ensure the practicality of the attack in the real world, the adversarial input should be correctly classified by humans as required by existing backdoor attacks~\cite{DBLP:conf/ndss/LiuMALZW018,DBLP:journals/corr/abs-2106-04690}.

\section{Clean Data-based Model Hijacking Attack}
\subsection{Overview}
Our goal is to generate a trigger derived from clean training data to compromise a target black-box model by activating its inherent natural backdoors. To achieve this, we first formalize the problem of generating the trigger as a multi-objective optimization task. However, the attacker cannot access the training and inference process of the target model. Auxiliary knowledge, such as the gradient information of the target model commonly employed in existing attacks, is currently unavailable. Hence, we train substitute models to emulate the behaviors of the target model to solve the optimization function. Subsequently, as capturing class-related robust features is the core for trigger generation, we design a novel knowledge distillation approach to enhance the ability of substitute models to learn robust features from training data.
In addition, we notice that most existing backdoor attacks generate triggers for specific models, which makes it difficult to work on other models. To further improve the generality of our attack, we design a sequential approach based on a meta-learning framework to craft triggers that share robust features across different models. The pipeline of CleanSheet is shown in Figure~\ref{fig:over}, which consists of generating a trigger based on substitute models and attacking the black-box models.

Algorithm \ref{algorithm-overview} outlines the workflow of CleanSheet. Lines 4–5 represent the generation of current adversarial inputs that will be fed into substitute models to optimize the triggers. Line 7 details the training of substitute models using knowledge distillation. Lines 9–12 delineate generating triggers based on a meta-learning approach.

\begin{figure}[t!]
\centering
\subfigure[Input]{\includegraphics[width=1.6cm, height=4.9cm]{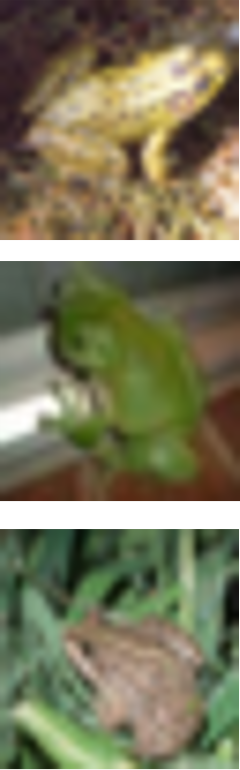}}
\subfigure[Epoch 3]{\includegraphics[width=1.6cm,height=4.9cm]{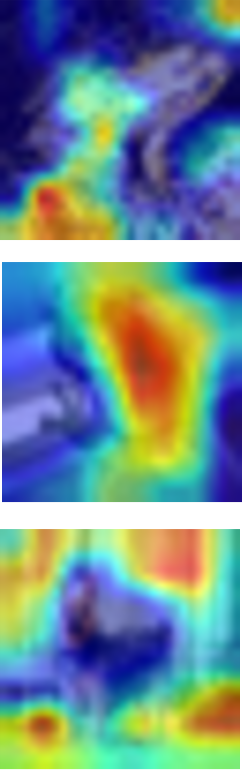}}
\subfigure[Epoch 5]{\includegraphics[width=1.6cm,height=4.9cm]{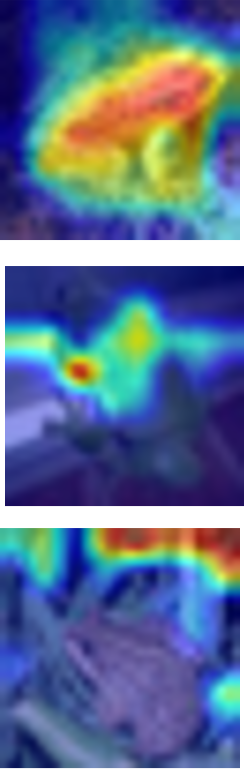}}
\subfigure[Epoch 50]{\includegraphics[width=1.6cm,height=4.9cm]{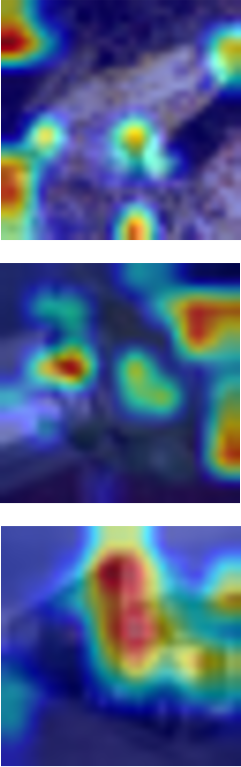}}
\subfigure[Epoch 150]{\includegraphics[width=1.6cm,height=4.9cm]{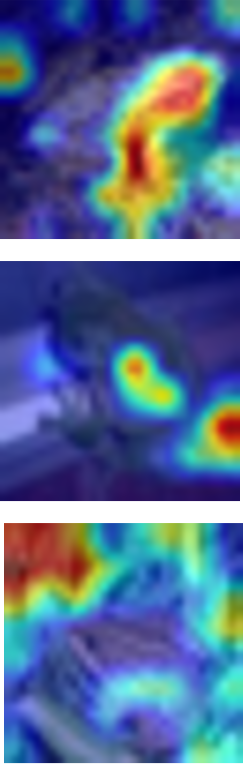}}
\vspace{-1em}
\caption{Attention maps on training epochs 3, 5, 50, and 150. The accuracy of the model on CIFAR-10 is 53.74\%, 64.49\%, 86.76\%, and 94.92\%, respectively.}
\label{fig:map}
\vspace{-1em}
\end{figure}

\subsection{Multi-objective Trigger Optimization}
To ensure that the adversarial inputs have the properties of hijacking usability and example invariance, we define the following multi-objective optimization problem:
\begin{equation}\label{eq:optimization_goal}
    \left\{\begin{matrix}
&\underset{\mathcal{T}(x)}{\arg\min} ~\mathbb{E}_{x\sim \mathcal{X}  } ~\ell(\mathcal{T}(x),y_t),
 \\
&\underset{\mathcal{T}(x)}{\arg\min} ~\mathbb{E}_{x\sim \mathcal{X}  }  ~\mathrm{D} (\mathcal{T}(x),x),
\end{matrix}\right.
\end{equation}
where $\mathcal{X}$ means the input space. The first objective is to transform the input $x$ into an adversarial input $\mathcal{T}(x)$ to achieve hijacking usability, accomplished by minimizing a loss function $\ell (\cdot)$, such as the cross-entropy function. The second objective aims to achieve the example invariance by minimizing the distance function $\mathrm{D}(\cdot)$, e.g., $l_p$-norm distance, between the adversarial and original inputs.

Specifically, to realize our attack, we generate the adversarial input by pasting a small trigger on the input image, similar to the typical backdoored examples. Hence, we formulate the input transformation function as follows:
\begin{equation}
    \begin{array}{c}
        \mathcal{T}(x)=(1-\mathbb{M})\odot x+\mathbb{M} \odot \Delta,
    \end{array}
\end{equation}
where $\mathbb{M}$ is a binary mask that indicates the location of the trigger. $\Delta$ indicates the value of the trigger, and $\odot$ symbolizes element-wise multiplication.  Notably, the shape of the trigger should remain inconspicuous.Therefore, the optimization objective of Eq. (\ref{eq:optimization_goal}) can be formalized as: 
\begin{equation}\label{eq:trigger}
    \underset{\mathbb{M},~\Delta }{\arg\min}~\mathbb{E}_{x\sim \mathcal{X}  } \left \{ \ell (f((1-\mathbb{M})\odot x+\mathbb{M} \odot \Delta ), y_t)+\lambda \cdot \mathrm{D}(\mathbb{M} )  \right \},
\end{equation}
where $\lambda$ represents the weight of the example invariance objective. The optimization process involves learning $\mathbb{M}$ and $\Delta$. We dynamically adjust the value of $\lambda $ to ensure that the ASR remains above 99\%.\footnote{Intuitively, this threshold might impact the ASR. However, our empirical analysis indicates that the ASRs on substitute models are insensitive to this parameter.}

A straightforward solution for solving Eq. (\ref{eq:trigger}) is to use the gradient descent algorithm. However, since the gradient information is inaccessible to the adversary in the black-box settings, we first need to establish a substitute model to estimate the gradients. Theoretically, if the substitute model has the same training objective and training data as the target model, they will likely share vulnerabilities, such as natural backdoors. Yet, our experimental results show that triggers generated based on the substitute models often fail to compromise the target black-box model, even though they can successfully attack the substitute model. We attribute this issue to two factors. Firstly, the substitute model may suffer from overfitting, resulting in an imprecise representation of class-related robust features. Secondly, different models may learn different robust features. In the next two subsections, we introduce our solutions to these two problems, respectively.

\subsection{Knowledge Distillation-based Learning Framework}
To prevent overfitting in substitute models, we introduce a new method based on knowledge distillation, termed competitive distillation. The key idea is to employ soft labels, which are probability vectors derived from an existing model, to guide the training of the substitute models. According to a previous study~\cite{DBLP:journals/corr/HintonVD15}, soft labels can improve model generalization on unknown data not contained in the training set. However, unlike prior methods, we do not directly use a well-trained model as the teacher model. Instead, we incorporate competitive learning by simultaneously training multiple substitute models and selecting the best-performing one to extract the soft labels. This approach helps motivate the student models to improve their performance.

We denote the group of substitute models as $ \mathcal{F} =\left \{ f_1,\dots , f_c \right \}$, where $c$ represents the number of the substitute models. 
At the start of training, a substitute model is randomly selected from $ \mathcal{F}$ to serve as the teacher model.
Then, we select the best-performing model $f_i$ from the group based on its validation accuracy, designating it as the teacher model after each epoch.
During the current epoch, the selected (teacher) model  $f_i$ with parameters $\theta_i$ exclusively learns knowledge from the dataset (known as hard-label data) by optimizing the following objective function:
\begin{align}
  \theta_i=  \underset{\theta_i}{\mathrm{argmin}}~\ell (x, y, \theta_i).
\end{align}

Once the teacher model is selected, we could extract its knowledge from the probability vectors. 
Given an input $x$, the embedded knowledge about $x$ could be encoded as follows:
\begin{equation}
    \begin{array}{c}
Z^{h }_{t}(x)=\left [ \frac{e^{z^{i}_{t}(x)/h } }{ {\textstyle \sum_{j=0}^{k-1}e^{z^{j}_{t}(x)/h}} }  \right ]_{i\in {0,\cdots,k-1}},
\end{array}
\end{equation} 
where $h$ represents the softmax temperature, $z_{t}$ symbolizes the logits derived from the teacher model, and $k$ represents the number of classes. Consequently, by integrating the knowledge of soft and hard labels, 
we can define the loss function for training other substitute (student) models  as follows:
\begin{equation}
\begin{aligned}
\centering
\mathcal{L}_{KD}= \alpha\cdot KL(Z_{j }(x),Z^h_{i}(x))&+ (1-\alpha)\cdot \ell (f_{j }(x), y),\\
     s.t. j\in g,&~ i\ne j,
    \end{aligned}
\end{equation}
where $\alpha $ serves as a hyperparameter controlling the extent of knowledge transferred from the teacher model, $h$ is the temperature parameter in the knowledge distillation process, and $g$ is the collection of student models. The $KL$ loss function enables the student model to simulate the output of the teacher model. The latter term uses hard labels in training.
In general, if we use a coefficient $tm_i$ to indicate whether a substitute model $f_i$ is the selected teacher model in the current epoch, the objective function of $f_i$ can be defined as:
\begin{equation}
\label{eq:su}
    \begin{aligned}
    \underset{\theta_i}{\mathrm{argmin}} ~\Big\{\sum_{j=1}^{c} tm_j\cdot  \alpha\cdot KL(Z_{i }(x),Z^h_{j}(x))
+
(1-\alpha )\cdot \ell(f_{i }(x), y)\Big\}\end{aligned}
\end{equation}

Moreover, we notice that increasing training epochs may lead to overfitting in substitute models. As illustrated in Figure~\ref{fig:map}, after the 5th epoch, the model already focuses more on the objects than the background in the image, so there is no need for hundreds of extra training epochs. 
This observation inspires us to take advantage of intermediate results during the training process instead of information from a fully trained model. 
We utilize the gradients obtained during these training processes to solve Eq.~(\ref{eq:trigger}). Specifically, we first train substitute models and then utilize the gradients calculated from these models to optimize the value and mask of the trigger according to the following objective function:
\begin{equation}
\label{eq.mt}
\begin{aligned}
    \Delta, \mathbb{M} = \underset{\Delta, \mathbb{M}}{\arg\min} ~ E_{x\sim \mathcal{X} }\left \{\sum_{i=0}^{c}  \ell(f_{i}(\mathcal{T}(x)),y_t)+\lambda \cdot D(\mathbb{M}) \right \}.
\end{aligned}
\end{equation}
Furthermore, as the performance of substitute models may vary during the training process, the values and masks should be optimized, allowing the trigger to adapt to substitute models of varying performance in each iteration. This approach also prevents the trigger from being exclusively effective on a single, fixed model, enabling the attacker to create an effective trigger across different models.

\begin{table*}[!t]
\centering\footnotesize
\caption{Performance of CleanSheet on CIFAR-10 and CIFAR-100.}
\label{TA:cifar10}
\begin{tabular}{ccccccc} 
\toprule
\multicolumn{7}{c}{CIFAR-10}                                                                                    \\ 
\midrule
Metric & ResNet-20 & VGG-11-BN & MobileNet V2 (0.5)  & ShuffleNet V2 0.5$\times$ & ShuffleNet V2 1.0$\times$ & RepVGG-A0  \\
CA(\%)  & 92.59    & 92.78     & 93.12               & 90.65               & 93.57               & 94.47       \\
ASR(\%) & 99.09    & 96.09     & 99.44               & 98.60               & 97.78               & 99.90       \\ 
\midrule
Metric & ResNet-44 & VGG-16-BN & MobileNet V2 (0.75) & ShuffleNet V2 1.5$\times$ & ShuffleNet V2 2.0$\times$ & RepVGG-A1  \\
CA(\%)  & 94.01    & 94.15     & 94.08               & 93.31               & 93.98               & 94.93       \\
ASR(\%) & 98.89    & 99.07     & 98.48               & 98.91               & 99.20               & 97.42          \\ 
\midrule
Metric & ResNet-56 & VGG-19-BN & MobileNet V2 (1.0)  & MobileNet V2 (1.4)  & VGG-13-BN           & RepVGG-A2  \\
CA(\%)  & 94.38    & 93.91     & 94.05               & 94.21               & 94.00                  & 95.27       \\
ASR(\%) & 99.19    & 99.29     & 99.23               & 99.65               & 99.15               & 97.80       \\
\midrule
\multicolumn{7}{c}{CIFAR-100}                                                                                    \\ 
\midrule
Metric  & ResNet-20 & VGG-11-BN & MobileNet V2 (0.5)  & ShuffleNet V2 0.5$\times$ & ShuffleNet V2 1.0$\times$ & RepVGG-A0 \\
CA(\%)   & 68.84    & 70.79     & 71.15               & 67.82               & 72.64               & 75.29      \\
ASR(\%)  & 98.64    & 90.38     & 99.26               & 89.84               & 97.49               & 99.14      \\
\midrule
Metric  & ResNet-32 & VGG-16-BN & MobileNet V2 (0.75) & ShuffleNet V2 1.5$\times$ & ShuffleNet V2 2.0$\times$ & RepVGG-A1 \\
CA(\%)   & 70.14    & 74.63     & 74.15               & 74.23               & 75.49               & 76.45      \\
ASR(\%)  & 98.22    & 98.86     & 96.55               & 98.63               & 93.82               & 99.81      \\
\midrule
Metric  & ResNet-44 & VGG-19-BN & MobileNet V2 (1.0)  & MobileNet V2 (1.4)  & VGG-13-BN           & RepVGG-A2 \\
CA(\%)   & 71.65    & 78.83     & 74.30               & 76.33               & 72.61               & 77.49      \\
ASR(\%)  & 99.45    & 99.84     & 98.82               & 99.67               & 97.95               & 99.42   \\   

\bottomrule
\end{tabular}
\\
\footnotesize{All the pre-trained target models for CIFAR-10 and CIFAR-100 are obtained from \url{https://github.com/chenyaofo/pytorch-cifar-models}.}

\end{table*}

\begin{table*}[!t]
\centering\footnotesize
\caption{Performance of CleanSheet on GTSRB and SVHN.}
\label{table:GS}
\resizebox{\linewidth}{!}{
\begin{tabular}{cccccccc} 
\toprule
\multicolumn{8}{c}{GTSRB}                                                                                                               \\ 
\midrule
Metric & ResNet-18   & ResNet-34   & ResNet-50   & MobileNet V2 (0.5)  & MobileNet V2 (0.75) & MobileNet V2 (1.4)  & MobileNet V2 (1.0)   \\
CA(\%)  & 98.22      & 97.51      & 97.62      & 97.26               & 98.13               & 97.77               & 97.93                \\
ASR(\%) & 91.61      & 90.70      & 91.31      & 93.92               & 93.63               & 98.03               & 90.46                \\ 
\midrule
Metric & RepVGG-A0 & RepVGG-A1 & RepVGG-A2 & ShuffleNet V2 0.5$\times$ & ShuffleNet V2 1.5$\times$ & ShuffleNet V2 1.0$\times$ & ShuffleNet V2 2.0$\times$  \\
CA(\%)  & 98.14      & 98.00      & 98.43      & 97.43               & 97.68               & 97.78               & 97.90                \\
ASR(\%) & 92.48      & 94.39      & 93.37      & 85.32               & 93.44               & 90.46               & 85.55                \\
 \midrule    
\multicolumn{8}{c}{SVHN}                                                                                                           \\
 \midrule    
Metric & ResNet-18   & ResNet-34   & ResNet-50   & MobileNet V2 (0.5)  & MobileNet V2 (0.75) & MobileNet V2 (1.4)  & MobileNet V2 (1.0)   \\
CA(\%)  & 96.10       & 96.37      & 96.44      & 92.69            & 95.58            & 95.53               & 95.60                 \\
ASR(\%) & 95.32      & 96.52      & 92.80      & 91.04               & 95.37               & 96.50               & 95.69                \\
\midrule
Metric & RepVGG-A0 & RepVGG-A1 & RepVGG-A2 & ShuffleNet V2 0.5$\times$ & ShuffleNet V2 1.5$\times$ & ShuffleNet V2 1.0$\times$ & ShuffleNet V2 2.0$\times$  \\
CA(\%)  & 96.55      & 96.49      & 96.65      & 95.19               & 95.83               & 95.23               & 95.74                \\
ASR(\%) & 97.07      & 96.79      & 96.46      & 92.63               & 93.64               & 96.01               & 93.10        \\       

\bottomrule
\end{tabular}}
\end{table*}

\subsection{Sequential Model-agnostic Meta-learning Framework}
We now present our strategy to solve the problem of excessive differences between the learned features of the substitute and target models, which can result in poor attack transferability.

Inspired by model-agnostic meta-learning (MAML), we propose a sequential model-agnostic meta-learning (SMAML) framework to generate the trigger with model-agnostic features. MAML is a model training framework that enables models to learn common features across multiple datasets (\eg, CIFAR-10 and SVHN)~\cite{DBLP:conf/icml/FinnAL17}. By analogy, we can regard the parameters of the trigger as those of a model, and different tasks as finding triggers for different models. Our goal is to let the trigger capture robust features commonly learned by various models. Therefore, we propose to apply MAML to the trigger generation process. Specifically, we use SMAML to optimize the mask $\mathbb{M}$ and value $\Delta$ of the trigger based on substitute models.

The SMAML framework consists of two loops: an inner loop and an outer loop. In the inner loop, triggers are generated for each substitute model, ensuring that these triggers encompass robust features of each model,  respectively. In the outer loop, the triggers produced in the inner loop are aggregated to formulate a new global trigger. Since it captures robust features from multiple models, it is effective for many models simultaneously. After the iterative optimization of the trigger, we finally obtain a universal trigger that can hijack various models.

\noindent\textbf{Inner Loop.}
For a substitute model $f_i$, we first create a temporary trigger $\Delta_i$ and a mask $\mathbb{M}_i$. Then, we use them to craft an adversarial input $\mathcal{T} (x)$, feed it into the next substitute model $f_{i+1}$, and minimize the following loss function to obtain a new trigger that could be effective on the model $f_{i+1}$:  
\begin{equation}
\label{eq.mt}
\begin{aligned}
   \Delta_{i+1}, \mathbb{M}_{i+1} &= \underset{\Delta_i, \mathbb{M}_i}{\arg\min} ~ E_{x\sim \mathcal{X} }\left \{ \ell(f_{i+1}(\mathcal{T}(x)),y_t)+\lambda \cdot D(\mathbb{M}_{i}) \right \},
   \\
& s.t.~ \mathcal{T} (x) =(1-\mathbb{M}_{i})\odot x+\mathbb{M}_{i}\odot \Delta_{i}.
\end{aligned}
\end{equation}

By carrying out this process across all substitute models sequentially, we can obtain a set of triggers $\left \{\Delta_{1},\Delta_{2},\cdots, \Delta_{c} \right \}$ and masks $\left \{\mathbb{M}_{1},\mathbb{M}_{2},\cdots, \mathbb{M}_{c} \right \}$ corresponding to models $\left \{f_{1},f_{2},\cdots, f_{c} \right \}$. This step aims to extract the robust features of each substitute model, thereby utilizing them to derive the common features across models.

\noindent\textbf{Outer Loop.}
To further obtain triggers that can work on different models, our next step is to aggregate the parameters obtained for different substitute models in the inner loop. We denote the trigger parameters for model $f_i$ as $\Delta_i$ and $\mathbb{M}_{i}$. We employ a simple averaging approach to calculate the global parameters for the next round of iterative optimization. This update strategy aims to find a trigger that can leverage the common features across multiple models.

\begin{table*}
\centering \footnotesize
\caption{Performance of CleanSheet on ImageNet.}
\label{TA:imagenet}
\begin{tabular}{cccccccccccc}
\toprule
\multicolumn{12}{c}{ImageNet}    \\                                                                                                                                    \midrule                                                                   
\multirow{2}{*}{Class} & \multirow{2}{*}{Metric} & \multicolumn{2}{c}{ResNet-18}           & \multicolumn{2}{c}{ResNet-34}             & \multicolumn{2}{c}{ResNet-50}    & \multicolumn{2}{c}{ResNet-101}   & \multicolumn{2}{c}{WRN-50-2}  \\
                       &                          & top-1 & top-5                          & top-1 & top-5                             & top-1 & top-5                   & top-1 & top-5                   & top-1 & top-5                          \\ 
\cmidrule{3-12}
\multirow{2}{*}{1000}  & CA(\%)                   & 76.52 & 92.62                          & 79.90  & 94.76                             & 83.80  & 95.54                   & 83.00    & 96.12                   & 83.74 & 96.38                          \\
                       & ASR(\%)                  & 58.32 & 85.40                           & 53.82 & 81.02                             & 45.64 & 78.84                   & 43.08 & 71.12                   & 32.14 & 63.66                          \\
\multirow{2}{*}{100}   & CA(\%)                   & 79.14 & 95.48                          & 82.02 & 96.64                             & 84.78 & 97.24                   & 84.52 & 97.60                    & 85.34 & 97.60                           \\
                       & ASR(\%)                  & 63.98 & 89.82                          & 58.16 & 85.80                              & 48.86 & 80.56                   & 46.84 & 78.12                   & 35.92 & 69.84                          \\ 
\midrule
\multirow{2}{*}{Class} & \multirow{2}{*}{Metric} & \multicolumn{2}{c}{WRN-101-2} & \multicolumn{2}{c}{VGG-11}                 & \multicolumn{2}{c}{VGG-11-BN}   & \multicolumn{2}{c}{VGG-16}       & \multicolumn{2}{c}{VGG-16-BN}          \\
                       &                          & top-1 & top-5                          & top-1 & top-5                             & top-1 & top-5                   & top-1 & top-5                   & top-1 & top-5                          \\ 
\cmidrule{3-12}
\multirow{2}{*}{1000}  & CA(\%)                   & 84.32 & 96.56                          & 77.24 & 93.06                             & 77.94 & 94.18                   & 79.20  & 94.24                   & 80.48 & 95.52                          \\
                       & ASR(\%)                  & 14.84 & 42.16                          & 59.50  & 81.94                             & 40.16 & 64.32                   & 50.38 & 74.92                   & 31.08 & 58.26                          \\
\multirow{2}{*}{100}   & CA(\%)                   & 85.8  & 97.70                           & 80.00    & 95.78                             & 80.48 & 96.28                   & 81.48 & 96.04                   & 82.64 & 97.16                          \\
                       & ASR(\%)                  & 21.84 & 54.38                          & 64.60  & 86.62                             & 43.56 & 69.18                   & 57.64 & 80.46                   & 38.76 & 66.82                          \\ 
\midrule
\multirow{2}{*}{Class} & \multirow{2}{*}{Metric} & \multicolumn{2}{c}{MobileNet V2}      & \multicolumn{2}{c}{ShuffleNet V2 1.0$\times$} & \multicolumn{2}{c}{DenseNet-121} & \multicolumn{2}{c}{DenseNet-169} & \multicolumn{2}{c}{DenseNet-201}        \\
                       &                          & top-1 & top-5                          & top-1 & top-5                             & top-1 & top-5                   & top-1 & top-5                   & top-1 & top-5                          \\ 
\cmidrule{3-12}
\multirow{2}{*}{1000}  & CA(\%)                   & 78.56 & 93.46                          & 77.12 & 92.30                              & 81.74 & 95.26                   & 82.14  & 95.72                    & 80.48 & 95.52                          \\
                       & ASR(\%)                  & 42.56 & 72.4                           & 27.46 & 57.92                             & 58.94 & 84.08                   & 42.26 & 69.56                    & 48.08 & 80.30                          \\
\multirow{2}{*}{100}   & CA(\%)                   & 78.56 & 93.46                          & 79.79 & 95.04                             & 84.22 & 97.18                   & 84.24 & 97.24                   & 82.64 & 97.16                          \\
                       & ASR(\%)                  & 50.08 & 79.98                          & 36.48 & 72.52                             & 61.58 & 87.40                   & 50.00    & 83.42                   & 38.76 & 66.82                          \\
\bottomrule
\end{tabular}
\\
\footnotesize{The triggers are optimized on substitute models trained on ImageNet-100. All the best clean models are from torchvision. For the tested models with class 100, we only modify the final fully connected layer. }
\vspace{-2mm}
\end{table*}

\section{Evaluation}\label{sec:evaluation}

\subsection{Experiment Setup}
\noindent\textbf{Datasets.} Our experiments utilize five widely used image datasets: (1) CIFAR-10~\cite{krizhevsky2009learning} is a small dataset for identifying pervasive objects, including 50,000 training and 10,000 test images from 10 classes.   (2) CIFAR-100~\cite{krizhevsky2009learning} is an extension of CIFAR-10. It has 100 categories with a total of 60,000 images, including 50,000 training images and 10,000 test images. (3) GTSRB~\cite{stallkamp2012man} involves 51,800 color road sign images, including 39,200 training images and 12,600 test images distributed in 43 categories. (4) SVHN~\cite{netzer2011reading} includes digit images from Google Street View House Number, which are categorized into 10 classes from digit 0 to digit 9. It consists of 73,257 training images and 26,032 test images.      (5) ImageNet~\cite{russakovsky2015imagenet} is a large visual object recognition dataset consisting of 1,281,167 training images, 50,000 validated images, and 100,000 test images, with 1000 categories.

\noindent\textbf{Metrics.} We use two metrics, \textit{Clean Accuracy} (CA) and \textit{Attack Success Rate} (ASR), for evaluation. CA measures the classification accuracy of the target model. ASR is the percentage of trigger-embedded testing instances that are recognized as the target class by the model. 

\noindent\textbf{Models.} 
For each dataset, we train the substitute models using four architectures to generate triggers, including ResNet-34~\cite{he2016deep}, ResNet-18~\cite{parkhi2015deep}, VGG-16~\cite{DBLP:journals/corr/abs-1802-02627}, and MobileNet V2~\cite{sandler2018mobilenetv2}. 
For model training, we use the SGD optimizer~\cite{goyal2017accurate} and set the learning rate as 0.2, momentum as 0.9, and weight decay as 0.0005.
For the hyperparameters of the knowledge distillation strategy, we follow the setting in the previous study~\cite{DBLP:journals/corr/HintonVD15}, where $h=1$ and $\alpha=0.5$. We set $\lambda =0.0001$ in Eq. (\ref{eq.mt}).

A total of 79 pre-trained models are used as the targets to evaluate our attack, including 18 models for CIFAR-10, 18 for CIFAR-100, 14 for GTSRB, 14 for SVHN, and 15 for ImageNet. All these models are obtained from GitHub or Torchvision without any modifications. Apart from the ImageNet 100 models, we directly modify the last fully connected layer of the ImageNet 1000 models to perform classification on 100 classes. For GTSRB and SVHN, we train 14 models under normal settings,  respectively.

\subsection{Attack Performance}
\noindent\textbf{Attack Success Rates.} In this subsection, for CIFAR-10, CIFAR-100, GTSRB, and SVHN, we assume the adversary has the complete training data of the target model.
The CAs and ASRs of these target models are presented in Tables~\ref{TA:cifar10},~\ref{table:GS}, and~\ref{TA:imagenet}. CleanSheet achieves an average ASR of 98.7\%, 97.5\%, 91.8\%, and 95.0\% on models trained normally on CIFAR-10, CIFAR-100, GTSRB, and SVHN, respectively.

For ImageNet, we assume the adversary only has about 10\% of the training data of the target model. Specifically, for the test model trained on ImageNet with 1,000 classes, we only use training data about the first 100 classes to train substitute models and then generate a trigger with respect to one of the first hundred classes. We still achieve an average ASR of 70.31\% (top-5). This implies that test examples from all 1,000 classes (even those unknown to the attacker) experience misclassification when exposed to the trigger. 
We find that the ASR on ImageNet is relatively lower than on other datasets. This can be attributed to the limited ability of the model to learn the features of the ImageNet dataset.
Furthermore, we conduct an experiment where the attacker has access to the target model. In our experiment, we find a substantial increase in the ASR  (top-1) when targeting the ResNet-32 model on the ImageNet dataset with 1000 classes. The ASR increases significantly from 53.82\% to 98.56\%. Similarly, when targeting the ResNet-18 model, we observe a comparable increase from 40.42\% to 98.56\%. 

Overall, CleanSheet can achieve comparable performance to previous backdoor attacks, \eg, an ASR of 89\% for clean-label backdoor attacks~\cite{DBLP:journals/corr/abs-1912-02771} and 97.29\% for traditional poisoning-based backdoor attacks~\cite{gong2022atteq} with a poisoning rate of 5\% on CIFAR-10, 
while eliminating the need for poisoned training data.

\noindent\textbf{Transferability.}
Since an adversary typically lacks access to the architecture information of the target model, we specifically evaluate the transferability of CleanSheet by using a group of triggers to attack models with different structures simultaneously. We attack 18 clean models with different depths and architectures. Table~\ref{TA:cifar10} shows that CleanSheet achieves an ASR of about 99\% for most models, including unknown models such as ShuffleNet and RepVGG. 
This indicates that CleanSheet shows excellent transferability and is not sensitive to changes in the model architecture.

\noindent\textbf{Attacking Speech Recognition Models.}
CleanSheet is also adaptable for attacking speech recognition models. We evaluate the performance of CleanSheet on models trained on the Google Speech Commands v2 dataset, which consists of 105,000 one-second-long audio files of 35 classes~\cite{DBLP:journals/corr/abs-1804-03209}. 
The target models include ECAPA-TDNN~\cite{DesplanquesTD20}, CNN~\cite{ArikKCHGFPC17}, ATT-RNN~\cite{deandrade2018neural}, and RNN~\cite{dong2018extending}, with CAs of 94.03\%, 90.42\%, 93.18\%, and 93.51\%, respectively. 
Across these models, CleanSheet achieves ASRs of 74.77\%, 71.96\%, 73.82\%, and 70.57\%, respectively, demonstrating its efficacy for attacking speech recognition models. 

Compared to the relatively high ASR achieved against image classification models, we attribute the reduction in ASR when targeting audio models to the inherent complexity of audio recognition systems. Audio systems are generally considered more complex than image classification systems, mainly due to the additional pre-processing step for extracting frequency features from the raw audio. The higher complexity presents challenges in extracting the robust features essential for successful attacks. Nonetheless, the performance of our attack against speech recognition models still shows its potential applicability.



\begin{figure*}[!t]
    \centering
    \includegraphics[width=0.98\linewidth]{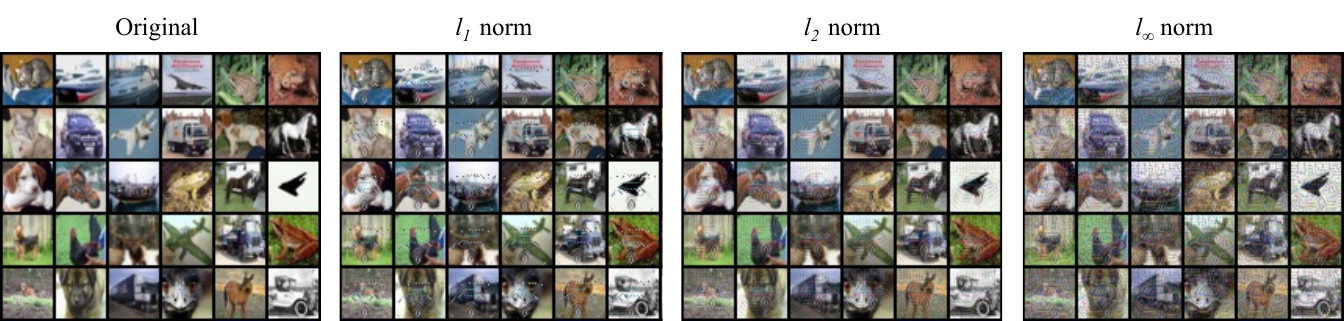}
    \vspace{-2mm}
    \caption{Adversarial inputs under different $l_p$ norm constraints and the corresponding nature instances.}
    \vspace{-5mm}
    \label{fig:lp}
\end{figure*}

\begin{table}[t]
\centering \footnotesize
\caption{Physical experiments of CleanSheet on CIFAR-10.}
\label{ta:physic}
\begin{tabular}{cccc} 
\toprule
Model & ASR(\%)  & Models & ASR(\%) \\ 
\midrule
MobileNet V2 (1.0) & 72.00  & ResNet-56 & 74.00  \\
MobileNet V2 (1.4) & 64.00 &ShuffleNet V2 1.5$\times$ & 52.00 \\
RepVGG-A0 & 74.00  &  ShuffleNet v2 2.0$\times$ & 55.00 \\
RepVGG-A1 & 67.00   & VGG-16-BN & 71.00  \\
ResNet-44 & 81.00 &  VGG-19-BN & 72.00  \\
\bottomrule
\end{tabular}
\end{table}

\noindent\textbf{Physical Attacks.}
Aside from the digital domain, we have also evaluated the effectiveness of CleanSheet in the physical world. For example, an attacker could print the trigger and affix it to traffic signs to deceive autonomous vehicles.

We randomly select 100 adversarial inputs for the CIFAR-10 dataset. Each sample is printed as a 4cm $\times$ 4cm image on white paper. We take a photo of each printed sample using an iPhone 12. Subsequently, we digitally crop the images to remove the edges of the white paper and resize them to 32 $\times$ 32 dimensions. Finally, we feed these images to the target models for classification. Table~\ref{ta:physic} demonstrates that CleanSheet is effective in the real world, achieving an average ASR of 68.2\% across 10 target models. 
We also observe a decrease in ASR compared with digital attacks. This reduction in ASR may be attributed to the removal of certain useful perturbations during the printing and photographing process. Therefore, we leave exploring and designing more robust physical  hijacking attacks for future work.

\noindent\textbf{Multi-trigger CleanSheet.}
Some prior attacks inject multiple backdoors into a target model, with each trigger corresponding to a different label~\cite{DBLP:journals/tdsc/XueHWL22}. These attacks are more adaptable and can potentially cause greater harm. We also explore how to extend CleanSheet to incorporate multiple triggers.

We generate several triggers for CIFAR-10 and calculate the ASR on different black-box models. The results are shown in Table~\ref{TA:multi} in Appendix~\ref{se:MT}. For CIFAR-10, the average ASRs of these triggers are close to 94\%. This confirms that CleanSheet can be used to generate multiple triggers against one model simultaneously. Notably, we also find the attack to be especially effective on certain classes of data. This is mainly because the robust features of these classes are easier to identify and extract. For example, the features of the class ``airplane'' are more distinct and easier to capture. 
\begin{table}[!t]
\centering\footnotesize
\caption{ASR(\%) of CleanSheet under the IID setting.  }
\label{overlap}
\resizebox{\linewidth}{!}{
\begin{tabular}{cccccc} 
\toprule
\multirow{2}{*}{\begin{tabular}[c]{@{}c@{}}Sub-dataset\\~of the attacker\end{tabular}} & \multicolumn{5}{c}{Sub-dataset of the user}                       \\ 
\cmidrule(lr){2-6}
                                                                             & {[}0, 0.5) & {[}0, 0.6) & {[}0, 0.7) & {[}0, 0.8) & {[}0, 0.9)  \\ 
\midrule
{[}0.1, 1)                                                                    & 94.70      & 90.15     & 96.39     & 96.04     & 92.73      \\
{[}0.2, 1)                                                                    & 94.61      & 93.92     & 94.79     & 97.59     & 95.78      \\
{[}0.3, 1)                                                                   & 89.77     & 82.41     & 87.64     & 91.21     & 91.33      \\
{[}0.4, 1)                                                                    & 91.64     & 86.57     & 85.94     & 86.63     & 88.97      \\
{[}0.5, 1)                                                                    & 93.88     & 90.86     & 90.34     & 94.45     & 91.05      \\
CA(\%)                                                                          & 91.53     & 92.13     & 92.99     & 93.40      & 93.89      \\

\multicolumn{6}{c}{CIFAR-10}                                                                                                               \\ 
\midrule
{[}0.1, 1)                                                                   & 90.99     & 90.45     & 91.35     & 98.81     & 98.19      \\
{[}0.2, 1)                                                                    & 80.87     & 78.94     & 82.29     & 97.34     & 95.25      \\
{[}0.3, 1)                                                                    & 89.28     & 89.64     & 95.21     & 98.83     & 98.46      \\
{[}0.4, 1)                                                                    & 51.26     & 39.11     & 64.56     & 95.80      & 94.39      \\
{[}0.5, 1)                                                                    & 75.81     & 64.96     & 82.88     & 99.04     & 98.71      \\
CA(\%)                                                                           & 69.19     & 70.22     & 72.06     & 73.57     & 74.53      \\
\multicolumn{6}{c}{CIFAR-100}                                                                                                              \\
\bottomrule
\end{tabular}}
\end{table}

\noindent \textbf{Comparison with UAP Attacks}.
To further highlight the superiority of CleanSheet, we compare it to UAP attacks. We reimplemented two classic UAP attacks~\cite{DBLP:journals/corr/Moosavi-Dezfooli16,hirano2020simple} and calculated their ASR on black-box models trained on CIFAR-10. Table~\ref{tab:uap} in Appendix shows that both attacks yielded unsatisfactory performance, achieving only an average ASR of 30.09\% and 37.98\%, respectively. These results are significantly lower than the ASR of 98.3\% achieved by CleanSheet.

\subsection{Distributions of Training Data} 
This subsection investigates how the amount of knowledge an attacker possesses about the training data affects the effectiveness of the attack.
Specifically, we consider three tiers of adversary capabilities. (1) The attacker has a portion of the training data of the target model. (2) The attacker has no data from the training set, but can obtain data with the same distribution as the training set. (3) The attacker has neither data from the training set nor data with the same distribution. We refer to the first two as the IID setting, wherein the adversary can directly initiate the attack. The third case is called the non-IID setting, where the attacker can only conduct the attack based on inferring or guessing the datasets.

\begin{table}[t]
\centering
\setlength{\tabcolsep}{3pt}
\caption{Performance of CleanSheet under the non-IID setting on CIFAR-10.}
\label{ta:noniid}
\resizebox{\linewidth}{!}{
\begin{tabular}{ccccc} 
\toprule
Model & CA(\%) & CA$_{non-iid}$(\%)  & ASR(\%) & ASR$_{non-iid}$(\%)   \\ 
\midrule
MobileNet V2 (0.5) & 93.12  & 77.80  (-15.32) & 99.44  & 95.67  (-3.77)     \\
MobileNet V2 (0.75)  & 94.08  & 80.88  (-13.20) & 98.48  & 98.26  (-0.22)    \\
MobileNet V2 (1.0)   & 94.05  & 79.68  (-14.37) & 99.23  & 91.30  (-7.93)    \\
MobileNet V2 (1.4)   & 94.21  & 84.40  (-9.81)  & 99.65  & 98.82  (-0.83)    \\
RepVGG-A0            & 94.47  & 80.54  (-13.93) & 99.90  & 96.17  (-3.73)    \\
RepVGG-A1            & 94.93  & 84.25  (-10.68) & 97.42  & 96.45  (-0.97)    \\
RepVGG-A2            & 95.27  & 82.50  (-12.77) & 97.80  & 99.85  (+2.05)    \\
ResNet-20            & 92.59  & 81.39  (-11.20) & 99.09  & 96.74  (-2.35)     \\
ResNet-44            & 94.01  & 83.62  (-10.39) & 98.89  & 98.87  (-0.02)    \\
ResNet-56            & 94.38  & 79.89  (-14.49) & 99.19  & 95.00  (-4.19)     \\
ShuffleNet V2 0.5$\times$   & 90.65  & 80.13  (-10.52) & 98.60  & 96.39  (-2.21)     \\
ShuffleNet V2 1.0$\times$   & 93.27  & 82.51  (-10.76) & 97.78  & 98.57  (+0.79)     \\
ShuffleNet V2 1.5$\times$   & 93.31  & 81.47  (-11.84) & 98.91  & 96.97  (-1.94)     \\
ShuffleNet V2 2.0$\times$   & 93.98  & 83.12  (-10.86) & 99.20  & 93.04  (-6.16)   \\
VGG-11-BN            & 92.78  & 81.05  (-11.73) & 96.09  & 77.21  (-18.88) \\
VGG-13-BN            & 94.00  & 83.76  (-10.24) & 99.15  & 94.10  (-5.05)   \\
VGG-16-BN            & 94.15  & 84.62  (-9.53)  & 99.07  & 98.95  (-0.12)    \\
VGG-19-BN            & 93.91  & 83.04  (-10.87) & 99.29  & 95.47  (-3.82)     \\      
\bottomrule
\end{tabular}}
\vspace{-4mm}
\end{table}

\noindent \textbf{IID Setting.} 
We first evaluate CleanSheet under the IID setting. We split the full original training set into 10 subsets, then use some of them to train the substitute model and the remaining to train the target model. For example, the dataset used by the target model is represented by [0, 0.5), indicating that it uses the first 5 subsets. If the dataset used by the substitute model is [0.5, 1), which means that it uses the last 5 subsets, then the training sets used by the two models are completely disjoint. By altering the subsets used to train these two models, we manipulate the overlap ratio of the training data. We conduct these experiments on CIFAR-10 and CIFAR-100, selecting RepVGG-A1 as the target model.

The results are presented in Table~\ref{overlap}. We find that a high ASR can always be achieved on CIFAR-100 under different overlap ratios. Thus, the attack can succeed as long as the adversary has some knowledge about the distribution of the training data. Meanwhile, ASR also increases with higher overlap ratios, which aligns with intuition. In particular, even when the datasets used to train the substitute and target models are completely disjoint, the ASR can still reach around 90\%. Moreover, we observe that target models with lower CAs are more vulnerable to our attack, mainly due to their weaker generalization ability and insufficient robustness.

\noindent\textbf{Non-IID Setting.}
We divide CIFAR-10 into two datasets using the Dirichlet distribution, one for the target model and another for the substitute model. This method is commonly used to simulate the non-IID setting in federated learning~\cite{DBLP:journals/corr/abs-1909-06335}.

The results are shown in Table~\ref{ta:noniid}. For comparison, we also provide the results when the target and substitute models use the same training data, representing the performance upper bound. As can be seen, compared to the IID setting, CA and ASR are lower under the non-IID setting. However, in most cases, the decrease in ASR is still small. Across the 18 target models, an average ASR as high as 95.43\% can still be achieved. We believe this is because examples from the same class but different distributions still share similar robust features. For example, images labeled as ``elephant'' always include robust features like ears, tusks, and trunks, irrespective of their distribution. Thus, as long as the attacker can capture the robust features of the target class, it can launch effective attacks regardless of the dataset it uses.

\begin{figure}[t]
    \centering
    \includegraphics[width=0.98\linewidth]{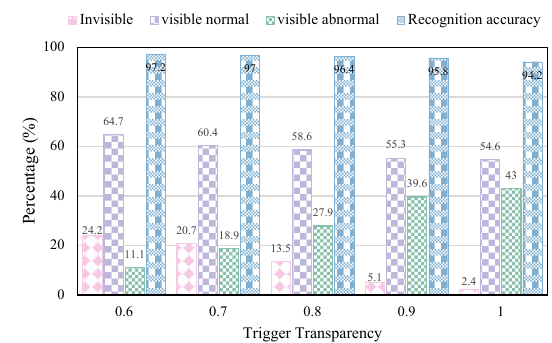}
    \vspace{-1mm}
    \caption{Evaluation results on human study of CleanSheet.}
    \vspace{-1mm}
    \label{fig:hum}
\end{figure}

\subsection{Imperceptibility}\label{se:lp}

\noindent\textbf{Constraint of Triggers.} As mentioned in Eq.~(\ref{eq:optimization_goal}), to satisfy the \emph{example invariance} property, we constrain the perturbation size of the trigger by introducing a distance function $D(\cdot)$. We considered three distance metrics: $l_1$ norm, $l_2$ norm, and $l_\infty$ norm, to limit the mask of the trigger. Figure~\ref{fig:lp} presents examples of adversarial inputs generated under these constraints, respectively. We observe that the generated triggers contain some visual features of the target class, which also validates that the triggers are generated based on class-related features. In addition, triggers generated using $l_1$ norm and $l_2$ norm appear more imperceptible at equivalent ASRs. Moreover, more visualization instances are provided in Figure~\ref{fig:ex} in Appendix~\ref{se:vs}.

\noindent\textbf{Transparency of Triggers.} 
We conduct further exploration on reducing the visibility of the trigger by adjusting its transparency. The results are shown in Table~\ref{tr:transparencies} in Appendix~\ref{se:tr}. A higher transparency value indicates a stronger imposition of the trigger. For CIFAR-10, even with a 0.6 trigger transparency (\ie, its transparency is only at 60\% of its original strength), CleanSheet still achieves an average ASR of 74.80\%. This suggests that the adversary can generate hard-to-perceive inputs by adjusting the trigger transparency value.
Besides, we note a trade-off between the attack performance and the invisibility. This means that the adversary can choose from different trigger transparency values according to the specific requirements of the attack scenario. 

\noindent\textbf{Human Perception.}
To assess if the adversarial inputs satisfy the \emph{example invariance} property, we conduct a survey with 20 volunteers, comprising 11 males and 9 females. The volunteers are recruited via online channels.\footnote{We have obtained ethical clearance from our institution. All volunteers were given a detailed explanation of the objectives of the experiment, and no sensitive or confidential information was requested.} For this experiment, we provide 50 adversarial inputs with trigger transparency levels varying from 0.6 to 1. We ask volunteers to identify the category of each evaluation sample and indicate it as normal, abnormal, or without visible triggers. 

As shown in Figure~\ref{fig:hum}, more than 95\% of the volunteers could accurately identify the true class of each adversarial input, indicating that the triggers did not impede human recognition. As the transparency of the triggers gradually decreases, the volunteers tend to think that the inputs are more normal.
Even when approximately 86\% of the volunteers noticed the triggers, around 60\% still perceived the images as normal, interpreting the trigger as a watermark or attributing it to low image quality, such as a photographic reflection. 
These results show that adversarial inputs generated by CleanSheet are also not easily noticed by humans.

\begin{table}[!t]
\centering \footnotesize
\setlength{\tabcolsep}{5pt}
 \caption{Ablation study.}
 \label{fig:ab}
 \resizebox{\linewidth}{!}{
\begin{tabular}{cccccc} 

\toprule
\multicolumn{6}{c}{CIFAR}                                                                                              \\ 
\midrule
\multirow{2}{*}{Case} & \multirow{2}{*}{Class} & ResNet & VGG & MobileNet & ShuffleNet  \\
                          &                         & top-1     & top-1      & top-1             & top-1                   \\ 
\midrule
\multirow{2}{*}{\textit{(w/o C\&S)}}   & 10                      & 53.22    & 42.82      & 46.89            & 24.63                  \\
                          & 100                     & 66.82    & 53.22     & 62.47           & 43.92                  \\ 

\multirow{2}{*}{\textit{(w/o CD)}}    & 10                      & 89.44    & 88.82     & 81.62            & 52.15                  \\
                          & 100                     & 92.59    & 92.96     & 92.39            & 84.10                   \\ 

\multirow{2}{*}{\textit{(w/o SMAML)}}   & 10                      & 80.27    & 82.59     & 79.54            & 59.08                  \\
                          & 100                     & 80.39    & 71.79     & 66.45            & 73.87                  \\ 

\multirow{2}{*}{\textit{Ours}}     & 10                      & \textbf{99.06}    & \textbf{99.37}    & \textbf{96.62}           & \textbf{93.41}                  \\
                          & 100                     & \textbf{99.45}    & \textbf{98.86}     & \textbf{99.26}            &\textbf{89.84}                  \\ 
\midrule
\multicolumn{6}{c}{ImageNet}                                                                                           \\ 
\midrule
\multirow{2}{*}{Case} & \multirow{2}{*}{Class} & ResNet& VGG     & MobileNet    & ShuffleNet  \\
                          &                         & top-5     & top-5      & top-5             & top-5                   \\ 
\midrule
\multirow{2}{*}{\textit{(w/o C\&S)}}   & 100                     & 64.60     & 56.64     & 50.18            & 47.92                  \\
                          & 1000                    & 53.36    & 46.82     & 36.16            & 30.52                  \\ 

\multirow{2}{*}{\textit{(w/o CD)}}     & 100                     & 63.64    & 84.16     & 69.24            & 40.10                   \\
                          & 1000                    & 53.14    & 79.86     & 59.09            & 19.24                  \\ 

\multirow{2}{*}{\textit{(w/o SMAML)}}    & 100                     & 53.82    & 57.64     & 59.32            & 36.48                  \\
                        & 1000                    & 48.86    & 54.04     & 45.68            & 36.30                   \\ 

\multirow{2}{*}{\textit{Ours}}     & 100                     & \textbf{80.56}    & \textbf{80.46}     & \textbf{79.98}            & \textbf{72.52}                  \\
                          & 1000                    & \textbf{78.84}    & \textbf{74.92}     & \textbf{72.40}             & \textbf{57.92}                  \\
\bottomrule
\end{tabular}}

\footnotesize{For CIFAR, we use ``Class'' to indicate the number of classes in CIFAR-10 or CIFAR-100. The target models are ResNet-44, VGG-16-BN, MobileNet V2 0.5, and ShuffleNet V2 0.5$\times$.
For ImageNet, we use ``Class'' to indicate ImageNet with the first 100 or 1000 classes. The target models are ResNet-50, VGG-16, MobileNet V2, and ShuffleNet V2 1.0$\times$. }
\vspace{-3mm}
\end{table}
\subsection{Ablation Study}
Competitive Distillation (CD) and Sequential Model-Agnostic Meta-Learning (SMAML) are two basic components for generating the triggers in our approach. To demonstrate their effectiveness, we conduct two ablation studies on CIFAR and ImageNet. We use the following terminology for clarity. ``\textit{Ours}''  refers to our approach utilizing CD and SMAML. The case ``\textit{w/o CD}'' refers to only using SMAML without CD. The case ``\textit{w/o SMAML}'' refers to only using CD without SMAML. The case ``\textit{w/o C\&S}'' refers to not using either of these two methods. The results are presented in Table~\ref {fig:ab}. 
We can see that the case ``\textit{w/o C\&S}'' still achieves a certain degree of ASR, which is significantly enhanced with the incorporation of CD. Using SMAML independently does not result in a significant improvement in ASR and may even cause a decrease in performance for some models.
However, using both methods together significantly improves ASR compared to using them alone.

\section{Defenses}
\subsection{Existing Defenses}\label{sec:existing_defenses}
\noindent\textbf{Model Pruning.} Prior studies have highlighted model pruning as an effective defense strategy against backdoor attacks~\cite{DBLP:conf/raid/0017DG18,DBLP:conf/iclr/LiLKLLM21}. This stems from the fact that clean examples and backdoored examples with triggers activate different parts of DNNs. Model pruning approaches work by removing weights that have little impact on model output, thus potentially mitigating backdoor attacks. The  CAs and ASRs of CleanSheet on some pruned models are shown in Table~\ref{tab:prune} in  Appendix~\ref{se:pr}. We can see that pruning affects both the CA of the model and the ASR of the attack. When the CA of the model decreases significantly, the ASR also drops. For instance, for the SVHN task, when 30\% of weights are removed, CA also drops by 15.1\%. For ShuffleNet V2 0.5$\times$, when the pruning rate increases from 0.25 to 0.3, ASR and CA decrease sharply, meaning that some important neurons are removed. However, for RepVGG-A0, MobileNet V2 (0.5), and ResNet-50, ASRs do not change significantly after removing some neurons. This observation suggests that CleanSheet exhibits a certain degree of resistance to model pruning, likely due to its activation of important neurons within the DNNs. 

\begin{table}[!t]
\centering
\setlength{\tabcolsep}{2pt}
\caption{Impact of fine-tuning and NAD on CleanSheet.}
\label{ta:ft}
\resizebox{\linewidth}{!}{
\begin{tabular}{ccccccc} 
\toprule
\multirow{2}{*}{Model} & \multicolumn{2}{c}{Vanilla~} & \multicolumn{2}{c}{Fine-tune} & \multicolumn{2}{c}{NAD}  \\ 
\cmidrule(lr){2-3}\cmidrule(r){4-5}\cmidrule(lr){6-7}
                        & CA(\%) & ASR(\%)             & CA(\%) & ASR(\%)              & CA(\%) & ASR(\%)         \\ 
\midrule
ResNet-20                & 92.59  & 99.09               & 96.33  & 90.86                & 89.95  & 96.8            \\
ResNet-32                & 93.53  & 99.28               & 92.34  & 91.85                & 90.33  & 94.95           \\
ResNet-44                & 94.01  & 98.89               & 94.87  & 91.71                & 91.10   & 94.78           \\
ResNet-56                & 94.38  & 99.19               & 87.85  & 91.84                & 91.32  & 91.49           \\
ShuffleNet V2 0.5$\times$     & 90.65  & 98.60                & 67.22  & 87.98                & 86.32  & 64.42           \\
ShuffleNet V2 1.0$\times$     & 93.57  & 97.78               & 82.27  & 90.17                & 89.30   & 83.97           \\
ShuffleNet V2 1.5$\times$     & 93.31  & 93.98               & 84.26  & 90.60                 & 90.21  & 84.11           \\
ShuffleNet V2 2.0$\times$     & 93.98  & 94.39               & 71.90   & 91.34                & 90.30   & 71.45           \\
\multicolumn{7}{c}{CIFAR-10}                                                                                      \\ 
\midrule
ResNet-20                & 68.84  & 98.64               & 64.46  & 94.23                & 63.06  & 92.14           \\
ResNet-32                & 70.14  & 98.22               & 66.54  & 95.61                & 63.45  & 93.01           \\
ResNet-44                & 71.65  & 99.45               & 67.51  & 95.84                & 65.61  & 92.46           \\
ResNet-56                & 72.61  & 97.95               & 68.79  & 85.92                & 65.45  & 79.01           \\
ShuffleNet V2 0.5$\times$     & 67.82  & 89.84               & 63.57  & 64.65                & 62.00     & 81.80            \\
ShuffleNet V2 1.0$\times$     & 72.64  & 91.28               & 67.92  & 86.36                & 66.21  & 75.29           \\
ShuffleNet V2 1.5$\times$     & 74.23  & 92.36               & 68.46  & 93.68                & 68.57  & 95.33           \\
ShuffleNet V2 2.0$\times$     & 75.49  & 92.95               & 70.85  & 68.86                & 68.21  & 74.15           \\
\multicolumn{7}{c}{CIFAR-100}                                                                                     \\
\bottomrule
\end{tabular}}

\end{table}

\noindent\textbf{Fine-tuning.} This is a straightforward yet effective method to defend against backdoor attacks by retraining the potentially backdoored model using clean data~\cite{DBLP:conf/raid/0017DG18}. This process could nullify the backdoor by altering the functionality of the model. 
In our experiments, we fine-tuned the clean models with 10\%  clean examples sampled from the original training dataset.
As shown in Table~\ref{ta:ft}, our attack maintains a high ASR after fine-tuning. This resilience is because CleanSheet leverages the normal classification capability of the target model instead of anomalous behaviors caused by poisoned data. Thus, it will not be invalidated by fine-tuning.

\noindent\textbf{Neural Attention Distillation (NAD).}
NAD~\cite{DBLP:conf/iclr/LiLKLLM21} is a classic defense method against backdoor attacks. Li~\etal observed that backdoored and normal neurons usually focus on different regions of the input image, motivating them to correct backdoored neurons through attention alignment~\cite{DBLP:conf/iclr/LiLKLLM21}. Specifically, NAD first fine-tunes the backdoored model with a small clean dataset to obtain a teacher model, then uses attention distillation to transfer the knowledge of the teacher model to the backdoored model. This process aligns backdoored neurons with normal ones. However, the ASR of CleanSheet only dropped by 11.05\% (still achieving an ASR of 85.32\%) after applying NAD. This result shows NAD also fails to defend against CleanSheet. This is because we utilize robust features instead of artificially crafted features for the attack. These robust features also exist in clean models.

\begin{table}[t]
\centering \footnotesize
\caption{Impact of Strip on CleanSheet.}
\label{ta:strip}
\resizebox{\linewidth}{!}{
\begin{tabular}{ccccc} 
\toprule
Model & $Mean_{clean}$ & $Std_{clean}$ & Threshold & $P_{escape}$(\%) \\ 
\midrule
MobileNet V2 (0.5) & 0.55 & 0.17 & 0.15 & 96.87 \\
MobileNet V2 (0.75) & 0.47 & 0.15 & 0.11 & 97.28 \\
MobileNet V2 (1.0) & 0.45 & 0.15 & 0.09 & 97.32 \\
MobileNet V2 (1.4) & 0.43 & 0.14 & 0.09 & 95.89  \\
RepVGG-A0 & 0.46 & 0.20 & -0.01 & 100.0 \\
RepVGG-A1 & 0.67 & 0.30 & -0.03 & 100.0 \\
RepVGG-A2 & 0.50 & 0.26 & -0.10 & 100.0 \\
ResNet-20 & 0.61 & 0.21 & 0.12 & 95.79  \\
ResNet-44 & 0.36 & 0.13 & 0.06 & 94.44 \\
ResNet-56 & 0.37 & 0.14 & 0.04 & 96.99 \\
ShuffleNet V2 0.5$\times$ & 0.78 & 0.24 & 0.20 & 99.89 \\
ShuffleNet V2 1.0$\times$ & 0.60 & 0.22 & 0.08 & 99.94 \\
ShuffleNet V2 1.5$\times$ & 0.54 & 0.19 & 0.08 & 97.53 \\
ShuffleNet v2 2.0$\times$ & 0.60 & 0.27 & -0.03 & 100.0  \\
VGG-11-BN & 0.38 & 0.14 & 0.05 & 99.98  \\
VGG-13-BN & 0.34 & 0.13 & 0.03 & 99.07 \\
VGG-16-BN & 0.23 & 0.09 & 0.01 & 98.99 \\
VGG-19-BN & 0.21 & 0.09 & 0.00 & 99.78  \\
\bottomrule
\end{tabular}}

\end{table}

\begin{table}[t]
\centering \footnotesize
\setlength{\tabcolsep}{2pt}
\caption{Impact of Beatrix on CleanSheet.}
\label{ta:beatrix}
\resizebox{\linewidth}{!}{
\begin{tabular}{cccccc} 
\toprule
Model & MAD & Detection & Models & MAD & Detection  \\ 
\midrule
MobileNet V2 (0.5) & 2.95 & \XSolidBrush & ResNet-56 & 1.39 & \XSolidBrush \\
MobileNet V2 (0.75) & 3.81 & \XSolidBrush & ShuffleNet V2 0.5$\times$ & 3.17 & \XSolidBrush \\
MobileNet V2 (1.0) & 1.73 & \XSolidBrush & ShuffleNet V2 1.0$\times$ & 3.50 & \XSolidBrush \\
MobileNet V2 (1.4) & 5.54 & \XSolidBrush & ShuffleNet V2 1.5$\times$ & 3.84 & \XSolidBrush \\
RepVGG-A0 & 1.03 & \XSolidBrush & ShuffleNet v2 2.0$\times$ & 4.35 & \XSolidBrush \\
RepVGG-A1 & 4.01 & \XSolidBrush & VGG-11-BN & 1.51 & \XSolidBrush \\
RepVGG-A2 & 4.16 & \XSolidBrush & VGG-13-BN & 2.69 & \XSolidBrush \\
ResNet-20 & 5.11 & \XSolidBrush & VGG-16-BN & 3.87 & \XSolidBrush \\
ResNet-44 & 4.00 & \XSolidBrush & VGG-19-BN & 5.03 & \XSolidBrush \\
\bottomrule
\end{tabular}}
\vspace{-3mm}
\end{table}

\noindent\textbf{Trigger Detection.}
Another type of defense is detecting malicious inputs during the inference phase. We tested two state-of-the-art detection methods, i.e., Strip~\cite{DBLP:conf/acsac/GaoXW0RN19} and Beatrix~\cite{DBLP:conf/ndss/MaWSXWX23}. Strip~\cite{DBLP:conf/acsac/GaoXW0RN19} detects triggers by observing changes in the output entropy when examples are combined with other clean data. Due to the strong correlation between triggers and target classes, examples with triggers exhibit significantly lower output entropy than normal examples. The detection results for Strip are presented in Table~\ref{ta:strip}, where $Mean_{clean}$ and $Std_{clean}$ represent the mean and standard deviation of the output entropy distribution estimated from the clean test set. $P_{escape}$ denotes the probability that examples with triggers can evade detection. The detection threshold is above the output entropy of 99\% of clean data. 
Beatrix~\cite{DBLP:conf/ndss/MaWSXWX23} quantifies high-order information and correlations between features by calculating the Gram matrix of intermediate model representations to determine if input examples are malicious. Clean examples and adversarial inputs can be effectively distinguished using this approach. As in the original setup, we compute the 1-9 order Gram matrix of the model representations as feature vectors and use the Median Absolute Deviation (MAD) to measure the deviation of adversarial inputs~\cite{leys2013detecting}. In our label detection process, we employ a constant value of $\eta=1.4826$ to calculate the anomaly index, as suggested in~\cite{leys2013detecting}. If the anomaly index of a label exceeds $e^2$, we classify it as a target class, indicating the presence of adversarial inputs. 

Both methods are tested on 18 models trained on CIFAR-10. Table~\ref{ta:beatrix} demonstrates that neither Strip nor Beatrix could successfully defend against CleanSheet. It exhibits an average escape probability of 98.32\% against Strip, while Beatrix fails to identify the target class of the attack in any of the models. This is because adversarial inputs generated by CleanSheet have robust features similar to those of normal examples regarding the target class. Therefore, the behavior and representation of adversarial inputs resemble those of normal examples, making the detection of CleanSheet challenging.

\begin{table}[t]
\caption{Performance of CleanSheet on robust models.}
\label{Ta:robust}
\centering
 \resizebox{\linewidth}{!}{
\begin{tabular}{ccccccc} 
\toprule
\multirow{2}{*}{Method} & \multicolumn{1}{c}{\multirow{2}{*}{Model}} & \multirow{2}{*}{CA(\%)} & \multirow{2}{*}{RA(\%)} & \multicolumn{3}{c}{ASR(\%)}  \\ 
\cmidrule(lr){5-7}
                        & \multicolumn{1}{c}{}                       &                         &                         & $l_1$ & $l_2$ & $l_\infty$   \\ 
\midrule
ComFact& WRN-70-16                           & 60.83                   & 49.46                   & 12.04 & 9.85  & 12.22        \\
 ComFact                       & WRN-70-16*                           & 69.15                   & 36.88                   & 26.46 & 29.50  & 24.70         \\ 
GenAug & WRN-70-16                           & 63.55                   & 34.64                   & 27.99 & 29.73 & 26.49        \\
GenAug & WRN-28-10                           & 62.41                   & 32.06                   & 16.90  & 20.89 & 21.11        \\
  GenAug   & WRN-34-10                           & 56.87                   & 28.50                    & 5.60   & 5.96  & 6.68         \\ 
SCORE   & WRN-28-10                           & 63.65                   & 31.08                   & 20.34 & 23.52 & 33.79        \\ 
HAT   & PreActResNet-18                            & 61.50                    & 28.88                   & 15.39 & 12.32 & 13.09        \\
\bottomrule
\end{tabular}}
\\
\footnotesize{
RA represents the best-known robust accuracy reported in their papers. $*$ means some extra data is used to train the model.
}

\end{table}

\subsection{Potential Countermeasures}
\noindent\textbf{Adversarial Training.}
As CleanSheet works by introducing triggers to the input, adversarial training naturally emerges as a potential defense mechanism that could decrease the sensitivity of the model to these triggers. Formally, for a given example $x$ with the label $y$, the goal of adversarial training is to establish a function: $f(x+\delta)=y$, while the attack seeks to find a $\delta$ such that $f(x+\delta)=t$, where $t\neq y$. As adversarial training and the attack operate towards opposing goals, adversarial training can theoretically offer a certain level of defense.

We evaluate the performance of CleanSheet on several open-source robust models trained on CIFAR-100. The adversarial training approach consists of two steps. First, untargeted adversarial examples are generated with the $l_\infty$-distance limited within $\frac{8}{255}$. Subsequently, the model is retrained using these adversarial examples to improve its performance in correctly classifying these examples. 
We evaluate 7 models trained with four different adversarial training strategies: ComFact~\cite{Gowal2020Uncovering}, GenAug~\cite{Rebuffi2021Fixing}, SCORE~\cite{Pang2022Robustness_WRN28_10}, and HAT~\cite{rade2021helper}. The results are provided in Table~\ref{Ta:robust}. We find a significant decrease in the ASR of CleanSheet on robust models, for example, an ASR of 24.7\% for the WRN-70-16 model. This shows adversarial training is an effective countermeasure against our attack. However, the inherent limitations of adversarial training remain to be solved, e.g., the trade-off between robustness and model accuracy, the high computational cost, etc.

\noindent\textbf{Protecting Training Data.}
Since realizing our attack requires the adversary to have knowledge about (a part of) the training data of the target model, the most effective and straightforward defense is to protect the information about the training data. During model training and usage, secure and authenticated data-sharing mechanisms should be established among all entities that can access the data, including all data owners and users. Specifically, more reliable data access control and management policies should be designed and adopted, such as only allowing authorized and trusted entities to use the data and adopting encrypted storage and transmission to prevent data leakage~\cite{mahawaga2022local}. In addition, it is vital to maintain the privacy of data source information. For example, if an attacker knows that an open-source dataset is used during training, this knowledge could facilitate the attack.

\section{Conclusion}
This paper presents CleanSheet, a new attack for hijacking DNNs by incorporating a discrete trigger into input samples. CleanSheet stands out by merging the strengths of conventional backdoor and adversarial example attacks, offering high success rates and robustness against black-box models while eliminating the need to access the training process or data of the target model. We introduce a knowledge distillation-based learning framework for training substitute models to generate triggers and a sequential model-agnostic meta-learning framework to enhance the generalizability of triggers.

Extensive experiments across 5 datasets, 79 normally trained models, 68 pruned models, and 39 defensive models validate the effectiveness of the proposed attack. CleanSheet can achieve impressive average ASRs of up to 98.3\% on the most popular image datasets. Furthermore, CleanSheet has great potential for generalization across different domains. Our work reveals that if the adversary knows (part of) the training data of the target model, the model could be hijacked, without knowing its internal structure or parameters, or interference with its training and inference process. Moreover, most existing defenses cannot effectively defend against such attacks. In light of these results, we emphasize the importance of protecting training data-related information to ensure the secure and reliable deployment of DNNs in the future.

\section*{Acknowledgments}
{
We thank the anonymous reviewers for their helpful and valuable feedback. This work was partially supported by National Key R\&D Program of China under Grant 2020AAA0107702, the NSFC under Grants U20B2049, U21B2018, 62302344, 62132011, 62161160337, 61822309,  61773310, U1736205, and 61802166, HK RGC under Grants R6021-20F and N\_CityU139/21, and Shaanxi Province Key Industry Innovation Program under Grant 2021ZDLGY01-02.
}
\bibliographystyle{plain}
\bibliography{a_css-sample}
\appendix

\begin{table}[!t]
\caption{Performance of
CleanSheet under different trigger transparencies. }
\label{tr:transparencies}
\centering
\setlength{\tabcolsep}{3pt}{
\resizebox{\linewidth}{!}{
\begin{tabular}{ccccccccccc} 
\toprule
\multicolumn{11}{c}{CIFAR-10}                                                                         \\
Transparency        & 0.1   & 0.2   & 0.3   & 0.4   & 0.5   & 0.6   & 0.7   & 0.8   & 0.9   & 1      \\ 
\midrule
VGG-11-BN           & 0.99  & 1.15  & 2.06  & 8.32  & 24.76 & 49.4  & 74.06 & 89.19 & 96.09 & 98.83  \\
VGG-13-BN           & 1.07  & 1.61  & 7.86  & 25.71 & 50.28 & 75.07 & 89.88 & 96.94 & 99.15 & 99.85  \\
VGG-16-BN           & 1.02  & 1.58  & 6.6   & 23.03 & 47.12 & 70.88 & 88.2  & 96.17 & 99.07 & 99.82  \\
VGG-19-BN           & 1.02  & 1.49  & 6.98  & 24.85 & 52.33 & 74.75 & 90.29 & 96.97 & 99.29 & 99.89  \\ 
RepVGG-A0        & 1.09  & 2.05  & 11.34 & 36.5  & 65.45 & 85.76 & 95.47 & 98.82 & 99.69 & 99.98  \\
RepVGG-A1          & 1.07  & 1.75  & 7.95  & 24.43 & 46.5  & 67.52 & 82.91 & 92.52 & 97.42 & 99.2   \\
RepVGG-A2          & 1.13  & 1.73  & 8.39  & 25.13 & 48.41 & 69.93 & 85.54 & 94.23 & 97.8  & 99.36  \\
ResNet-20            & 1.09  & 2.27  & 10.79 & 29.83 & 54.93 & 76.51 & 90.21 & 96.69 & 99.09 & 99.87  \\
ResNet-32            & 1     & 2.08  & 11.14 & 33.92 & 60.85 & 82.18 & 92.58 & 97.63 & 99.28 & 99.78  \\
ResNet-44            & 1.14  & 2.61  & 10.46 & 27.69 & 51.63 & 74.4  & 88.67 & 96.28 & 98.89 & 99.78  \\
ResNet-56            & 1.08  & 2.25  & 10.93 & 30.47 & 56.56 & 78.51 & 91.6  & 97.29 & 99.19 & 99.83  \\
ShuffleNet V2 0.5$\times$ & 1.08  & 2.06  & 8.85  & 26.2  & 51.02 & 73.87 & 88.73 & 95.84 & 98.69 & 99.55  \\
ShuffleNet V2 1.0$\times$ & 1.08  & 1.82  & 7.53  & 22.51 & 44.33 & 66.12 & 82.97 & 93.07 & 97.78 & 99.4   \\
ShuffleNet V2 1.5$\times$ & 1.15  & 1.82  & 7.55  & 23.88 & 48.25 & 71.99 & 88.07 & 96.02 & 98.91 & 99.69  \\
ShuffleNet V2 2.0$\times$ & 1.11  & 1.98  & 8.45  & 25.84 & 51.35 & 74.95 & 89.55 & 96.6  & 99.2  & 99.81  \\
MobileNet V2 (0.5)  & 1     & 2.26  & 10.53 & 30.8  & 56.79 & 79.37 & 92.14 & 97.73 & 99.44 & 99.93  \\
MobileNet V2 (0.75) & 1.09  & 2.53  & 10.55 & 27.27 & 48.41 & 70.33 & 86.49 & 95.12 & 98.48 & 99.61  \\
MobileNet V2 (1.0)  & 1.03  & 2.71  & 12.53 & 33.77 & 57.73 & 78.6  & 91.33 & 97.39 & 99.23 & 99.8   \\
MobileNet V2 (1.4)  & 1.11  & 2.67  & 15.08 & 41.28 & 67.95 & 86.41 & 95.29 & 98.42 & 99.65 & 99.94  \\

\midrule
\multicolumn{11}{c}{CIFAR-100}                                                                        \\
Transparency        & 0.1   & 0.2   & 0.3   & 0.4   & 0.5   & 0.6   & 0.7   & 0.8   & 0.9   & 1      \\ 
\midrule
VGG-11-BN           & 9.92  & 9.91  & 9.96  & 11.21 & 15.21 & 24.26 & 38.81 & 57.8  & 75.34 & 87.63  \\
VGG-13-BN           & 9.98  & 10.05 & 11.93 & 20.32 & 37.5  & 60.46 & 80.38 & 92.32 & 97.46 & 99.39  \\
VGG-16-BN           & 10.05 & 10.16 & 13.39 & 23.92 & 43.76 & 67.01 & 86.05 & 95.12 & 98.68 & 99.75  \\
VGG-19-BN           & 10.13 & 10.38 & 13.74 & 25.07 & 44.4  & 68.59 & 86.21 & 95.46 & 98.7  & 99.73  \\
RepVGG-A0        & 10.05 & 10.72 & 15.52 & 27.13 & 45.79 & 66.74 & 83.02 & 93.12 & 97.44 & 99.22  \\
RepVGG-A1          & 10.04 & 10.54 & 14.18 & 23.59 & 39.37 & 58.88 & 76.21 & 88.89 & 95.74 & 98.52  \\
RepVGG-A2          & 10.03 & 10.69 & 15.24 & 28.78 & 50.07 & 71.35 & 86.98 & 94.48 & 98.07 & 99.33  \\
ResNet-20            & 10    & 10.69 & 15.38 & 28.55 & 49.14 & 71.43 & 86.52 & 94.96 & 98.28 & 99.44  \\
ResNet-32            & 9.99  & 10.78 & 15.64 & 29.41 & 50.84 & 72.81 & 88.12 & 95.73 & 98.81 & 99.64  \\
ResNet-44            & 10.23 & 10.9  & 16.35 & 30.28 & 49.69 & 70.36 & 85.17 & 93.45 & 97.75 & 99.15  \\
ResNet-56            & 10.03 & 11.32 & 17.23 & 30.81 & 50.87 & 71.83 & 86.78 & 94.83 & 98.26 & 99.48  \\
ShuffleNet V2 0.5$\times$ & 10.06 & 10.04 & 11.08 & 15.26 & 24.86 & 40.88 & 59.71 & 76.12 & 86.99 & 93.48  \\
ShuffleNet V2 1.0$\times$ & 9.94  & 10.09 & 11.61 & 16.91 & 28.32 & 45.16 & 63.75 & 79.09 & 89.48 & 94.95  \\
ShuffleNet V2 1.5$\times$ & 10.12 & 10.53 & 13.76 & 22.98 & 38.08 & 56.47 & 72.9  & 84.74 & 91.71 & 95.89  \\
ShuffleNet V2 2.0$\times$ & 9.98  & 10.16 & 11.95 & 18.18 & 30.99 & 48.84 & 67.14 & 82.27 & 91.69 & 96.76  \\
MobileNet V2 (0.5)  & 9.93  & 10.65 & 15.76 & 27.44 & 46.37 & 67.62 & 84.26 & 93.46 & 97.86 & 99.27  \\
MobileNet V2 (0.75) & 9.89  & 10.48 & 14.73 & 26.32 & 45.44 & 66.54 & 83.2  & 93.1  & 97.54 & 99.2   \\
MobileNet V2 (1.0)  & 9.97  & 10.37 & 13.45 & 22.47 & 38.4  & 58.1  & 75.95 & 88.28 & 95.31 & 98.39  \\
MobileNet V2 (1.4)  & 10.07 & 10.44 & 13.74 & 23.52 & 40.81 & 63.05 & 82.07 & 92.64 & 97.43 & 99.17  \\

\midrule
\multicolumn{11}{c}{GTSRB}                                                                            \\
Transparency
                       & 0.10 & 0.20  & 0.30  & 0.40  & 0.50  & 0.60  & 0.70  & 0.80  & 0.90  & 1.00   \\ 
\midrule

ResNet-18               & 7.98 & 18.80 & 32.26 & 46.04 & 60.23 & 70.84 & 79.29 & 86.58 & 91.61 & 95.08  \\
ResNet-34               & 8.20 & 18.84 & 32.87 & 48.14 & 61.58 & 72.63 & 81.34 & 86.94 & 90.70 & 93.68  \\
ResNet-50               & 8.29 & 17.52 & 30.98 & 45.71 & 58.80 & 70.32 & 78.92 & 86.09 & 91.31 & 95.11  \\
RepVGG-A0             & 6.84 & 15.41 & 28.22 & 41.46 & 54.00 & 66.13 & 77.17 & 86.17 & 92.48 & 96.78  \\
RepVGG-A1             & 7.04 & 16.02 & 30.66 & 46.93 & 60.70 & 71.53 & 80.59 & 88.67 & 94.39 & 97.49  \\
RepVGG-A2             & 7.39 & 18.43 & 32.42 & 47.47 & 61.14 & 72.58 & 82.40 & 88.84 & 93.37 & 96.04  \\
MobileNet V2 (0.5)     & 6.98 & 16.34 & 30.69 & 44.50 & 56.73 & 69.18 & 80.79 & 88.85 & 93.92 & 96.64  \\
MobileNet V2 (0.75)    & 7.43 & 17.92 & 32.84 & 46.86 & 59.28 & 70.19 & 79.85 & 87.90 & 93.63 & 97.85  \\
MobileNet V2 (1.4)     & 7.74 & 18.44 & 34.13 & 48.94 & 62.62 & 75.07 & 85.81 & 93.30 & 98.03 & 99.52  \\
MobileNet V2 (1.0)     & 6.74 & 15.34 & 29.14 & 41.82 & 54.04 & 65.00 & 75.14 & 83.94 & 90.46 & 94.71  \\
ShuffleNet V2 0.5$\times$    & 8.46 & 19.49 & 31.70 & 43.53 & 54.84 & 64.62 & 73.08 & 79.88 & 85.32 & 89.77  \\
ShuffleNet V2 1.5$\times$    & 7.66 & 16.69 & 29.80 & 42.00 & 55.12 & 67.31 & 78.32 & 86.32 & 93.44 & 96.98  \\
ShuffleNet V2 1.0$\times$    & 7.41 & 18.55 & 31.76 & 44.48 & 57.30 & 68.69 & 77.98 & 85.53 & 90.46 & 93.94  \\
ShuffleNet V2 2.0$\times$    & 7.70 & 17.85 & 31.64 & 43.78 & 55.47 & 65.84 & 74.71 & 80.82 & 85.55 & 88.88  \\
\midrule
\multicolumn{11}{c}{SVHN}                                                                            \\
Transparency               & 0.1   & 0.2   & 0.3   & 0.4   & 0.5   & 0.6   & 0.7   & 0.8   & 0.9   & 1      \\ 
\midrule
ResNet-18            & 20.39 & 22.70 & 29.00 & 40.27 & 53.02 & 65.70 & 76.57 & 85.08 & 91.28 & 95.32  \\
ResNet-34            & 20.21 & 22.69 & 30.32 & 42.76 & 56.84 & 69.62 & 80.07 & 87.72 & 93.09 & 96.52  \\
ResNet-50            & 20.00 & 21.65 & 27.08 & 37.14 & 49.45 & 61.75 & 72.81 & 81.52 & 88.10 & 92.80  \\
RepVGG-A0          & 20.16 & 22.25 & 29.40 & 42.64 & 57.23 & 71.03 & 81.90 & 89.47 & 94.31 & 97.07  \\
RepVGG-A1          & 20.01 & 22.30 & 29.43 & 42.06 & 56.68 & 69.78 & 80.71 & 88.46 & 93.72 & 96.79  \\
RepVGG-A2          & 20.11 & 22.75 & 30.39 & 43.25 & 57.73 & 70.47 & 80.96 & 88.18 & 93.37 & 96.46  \\
MobileNet V2 (0.5)  & 20.55 & 23.62 & 30.44 & 40.45 & 51.79 & 62.49 & 72.36 & 80.52 & 86.63 & 91.04  \\
MobileNet V2 (0.75) & 20.09 & 22.48 & 30.08 & 41.94 & 55.83 & 68.29 & 78.76 & 86.70 & 92.04 & 95.37  \\
MobileNet V2 (1.4)  & 19.88 & 22.86 & 31.09 & 44.16 & 58.13 & 70.90 & 80.95 & 88.66 & 93.44 & 96.50  \\
MobileNet V2 (1.0)  & 20.23 & 22.75 & 30.38 & 42.01 & 55.23 & 67.85 & 78.35 & 86.37 & 91.96 & 95.69  \\
ShuffleNet V2 0.5$\times$ & 20.18 & 22.23 & 28.58 & 39.02 & 51.14 & 63.35 & 73.73 & 82.19 & 88.29 & 92.63  \\
ShuffleNet V2 1.5$\times$ & 20.29 & 23.29 & 31.09 & 42.64 & 54.77 & 66.14 & 76.42 & 84.28 & 89.87 & 93.64  \\
ShuffleNet V2 1.0$\times$ & 20.40 & 23.08 & 30.93 & 43.26 & 57.24 & 69.70 & 79.97 & 87.48 & 92.60 & 96.01  \\
ShuffleNet V2 2.0$\times$ & 20.21 & 22.63 & 29.34 & 40.02 & 51.94 & 63.58 & 73.84 & 82.38 & 88.67 & 93.10  \\
\bottomrule
\end{tabular}}}
\end{table}

\begin{table*}[!t]
\centering \footnotesize
\caption{ASR(\%) of UAP attacks on black-box models.}
\label{tab:uap}

\begin{tabular}{ccccccccc}
\toprule
\multirow{2}{*}{\diagbox[dir=NW]{Test model}{Generate model}} & \multicolumn{4}{c}{UAP (target)~\cite{DBLP:journals/corr/Moosavi-Dezfooli16}}              & \multicolumn{4}{c}{Iterative~\cite{hirano2020simple}}                \\
\cmidrule(lr){2-5} \cmidrule(lr){6-9}
 & ResNet34 & ResNet18 & VGG16   & MobileNet V2 & ResNet34 & ResNet18 & VGG16   & MobileNet V2 \\
 \midrule
White-box                 & 95.35  & 93.46  & 86.49 & 82.85      & 97.93  & 97.13  & 78.20 & 93.73      \\
MobileNet V2 (0.5)        & 14.90  & 12.32  & 62.31 & 39.82      & 17.13  & 26.63  & 61.47 & 55.88      \\
MobileNet V2 (0.75)       & 14.64  & 12.36  & 46.58 & 44.45      & 16.93  & 23.35  & 53.95 & 52.98      \\
MobileNet V2 (1.0)        & 12.05  & 8.73   & 52.28 & 30.96      & 17.49  & 16.70  & 51.67 & 53.56      \\
MobileNet V2 (1.4)        & 8.32   & 7.87   & 47.77 & 38.36      & 12.27  & 15.74  & 49.16 & 66.30      \\
RepVGG-A0                 & 14.66  & 15.21  & 65.53 & 39.95      & 22.55  & 31.06  & 58.11 & 56.09      \\
RepVGG-A1                 & 16.12  & 15.02  & 59.91 & 29.84      & 25.71  & 31.38  & 52.35 & 38.07      \\
RepVGG-A2                 & 6.32   & 3.17   & 47.22 & 18.65      & 8.06   & 7.59   & 53.62 & 43.41      \\
ResNet-20                 & 32.60  & 19.08  & 57.84 & 40.28      & 39.85  & 31.15  & 60.36 & 64.11      \\
ResNet-44                 & 9.15   & 9.22   & 59.38 & 27.82      & 14.33  & 22.08  & 59.25 & 55.96      \\
ResNet-56                 & 13.80  & 9.80   & 57.71 & 35.63      & 20.60  & 18.06  & 65.68 & 56.27      \\
ShuffleNet V2 0.5$\times$        & 12.15  & 11.33  & 30.52 & 28.44      & 12.07  & 15.16  & 34.18 & 48.12      \\
ShuffleNet V2 1.0$\times$        & 13.20  & 15.99  & 35.36 & 18.76      & 23.97  & 24.13  & 37.82 & 36.65      \\
ShuffleNet V2 1.5$\times$        & 74.69  & 77.79  & 65.18 & 56.57      & 84.00  & 86.76  & 52.28 & 60.38      \\
ShuffleNet V2 2.0$\times$        & 8.71   & 8.00   & 37.75 & 17.62      & 14.30  & 15.18  & 40.95 & 25.70      \\
VGG-11-BN                 & 9.71   & 9.86   & 15.32 & 10.95      & 10.04  & 10.27  & 15.53 & 12.06      \\
VGG-13-BN                 & 12.34  & 11.86  & 59.38 & 16.00      & 14.37  & 18.10  & 53.21 & 28.92      \\
VGG-16-BN                 & 22.81  & 16.29  & 69.16 & 23.45      & 29.53  & 34.61  & 62.44 & 41.92      \\
VGG-19-BN                 & 10.08  & 9.97   & 58.92 & 21.60      & 12.95  & 13.65  & 57.19 & 31.93     \\
\bottomrule
\end{tabular}
\vspace{-1mm}
\end{table*}

\begin{figure*}[t]
    \centering
    \includegraphics[width=0.9\linewidth]{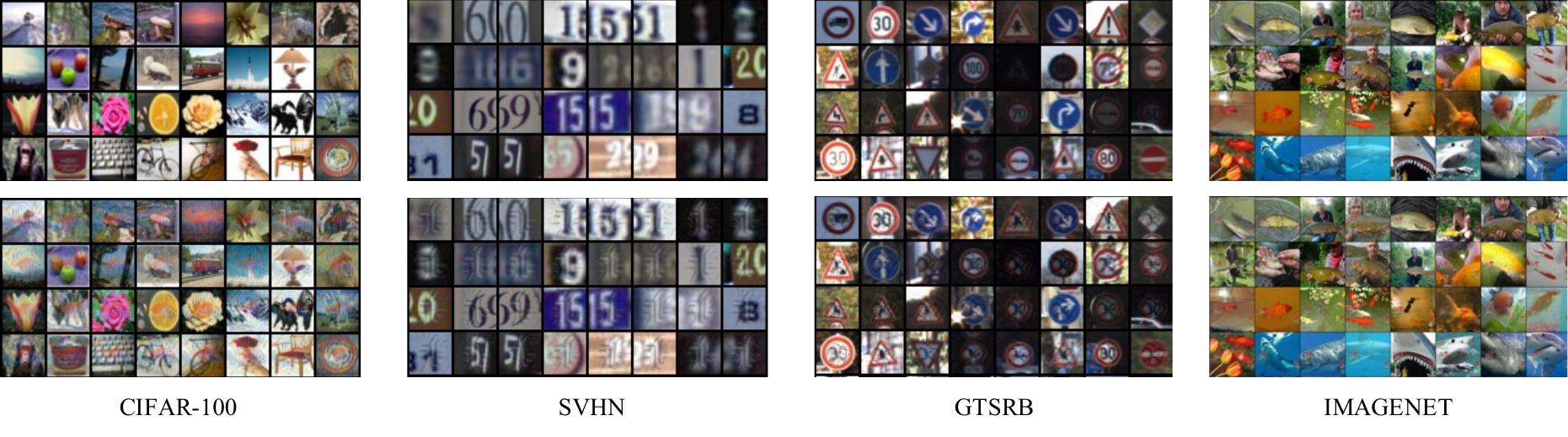}
\vspace{-1em}
    \caption{Adversarial inputs and clean examples. The top line is from clean examples, and the bottom is from adversarial inputs.}
\vspace{-1em}
    \label{fig:ex}
\end{figure*}


\subsection{Performance under Different $l_p$ Norms and Trigger Transparencies}\label{se:tr} 
The detailed performance of CleanSheet, applied with various $l_p$ norms on CIFAR-10 and CIFAR-100, is presented in Table~\ref{lp:cifar}. The triggers are close to the same size. We can observe that the $l_1$ norm performs worse than the other two norms in terms of ASR. Even so, the clean models will still be attacked.
Table~\ref{tr:transparencies} presents the ASRs of CleanSheet under different trigger transparency levels on CIFAR-10. We can observe that CleanSheet is still effective against clean models with varied architectures, even when the trigger is reduced to only 60\% of its original size.  

\subsection{Visualization Results}\label{se:vs}
\begin{figure}[h]
    \centering
    \includegraphics[width=\linewidth]{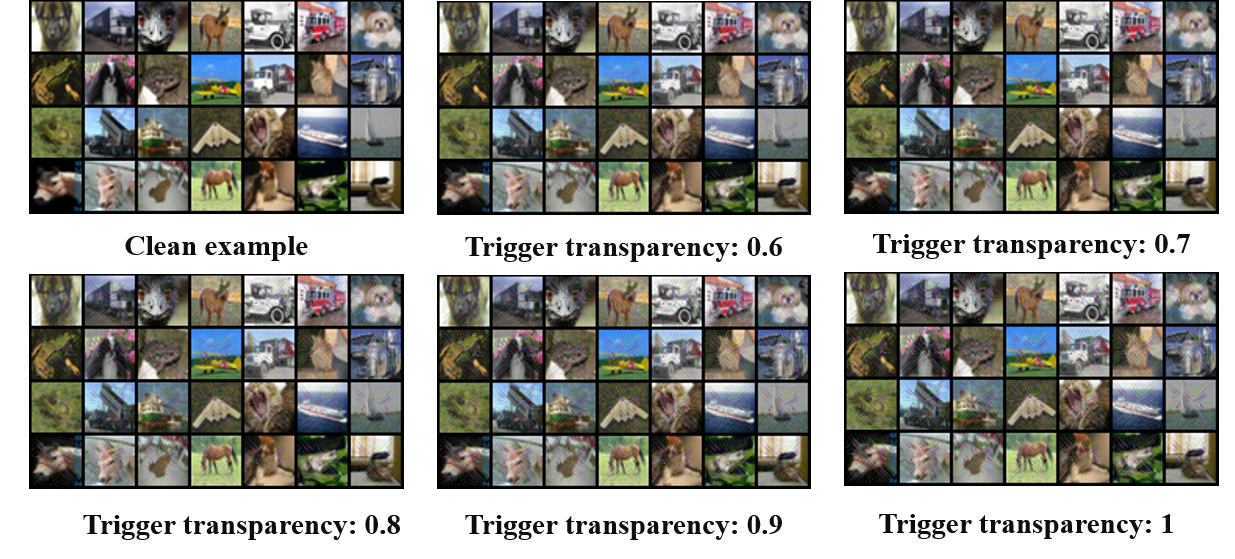}
\vspace{-2em}
    \caption{The adversarial inputs under different trigger transparencies.}
    \label{fig:tra}
\end{figure}

Figure~\ref{fig:ex} and Figure~\ref{fig:tra} visually presents adversarial inputs containing the trigger and the corresponding original examples.
We highlight that, though the trigger can be observed in the inference phase, we do not need to add any malicious examples in the training phase. Therefore, CleanSheet can escape manual checks, usually in the data collection phase.

\begin{table}[t]
\centering \footnotesize
\caption{Physical experiments of CleanSheet on CIFAR-10.}
\label{ta:physic}
\setlength{\tabcolsep}{3pt}{
\resizebox{0.95\linewidth}{!}{
\begin{tabular}{cccc} 
\toprule
Models & ASR(\%)  & Models & ASR(\%) \\ 
\midrule
MobileNet V2 (0.5) & 72  & ResNet-56 & 74 \\
MobileNet V2 (0.75) & 67  & ShuffleNet V2 0.5$\times$ & 41  \\
MobileNet V2 (1.0) & 72  & ShuffleNet V2 1.0$\times$ & 53  \\
MobileNet V2 (1.4) & 64  & ShuffleNet V2 1.5$\times$ & 52  \\
RepVGG-A0 & 74 & ShuffleNet v2 2.0$\times$ & 55  \\
RepVGG-A1 & 67  & VGG-11-BN & 21  \\
RepVGG-A2 & 70  & VGG-13-BN & 66  \\
ResNet-20 & 71  & VGG-16-BN & 71  \\
ResNet-44 & 81  & VGG-19-BN & 72  \\
\bottomrule
\end{tabular}}}
\vspace{-4mm}
\end{table}

\begin{table}[!t]
\centering \footnotesize
\caption{Performance of CleanSheet with different $l_p$ norms.}
\label{lp:cifar}
\setlength{\tabcolsep}{2.5pt}{
\resizebox{0.9\linewidth}{!}{
\begin{tabular}{ccccccc} 
\toprule
\multirow{2}{*}{Models} & \multicolumn{3}{c}{CIFAR-10~} & \multicolumn{3}{c}{CIFAR-100}  \\ 
\cmidrule(lr){2-4} \cmidrule(lr){5-7}
                        & $l_1$     & $l_2$    & $l_{\infty }$            & $l_1$     & $l_2$    & $l_{\infty }$            \\ 
\midrule
VGG-11-BN               & 62.65 & 68.34 & 80.39        & 62.16 & 83.07 & 89            \\
VGG-13-BN               & 91.33 & 98.69 & 99.01        & 81.14 & 96.9  & 97.43         \\
VGG-16-BN               & 96    & 99.35 & 99.76        & 87.89 & 96.4  & 98.2          \\
VGG-19-BN               & 95.79 & 99.45 & 99.72        & 97.04 & 97.34 & 98.96         \\
RepVGG-A0            & 86.96 & 99.35 & 99.85        & 81.39 & 96.11 & 99.19         \\
RepVGG-A1              & 75.86 & 98.19 & 99.73        & 95.44 & 98.14 & 98.64         \\
RepVGG-A2              & 85.89 & 99.03 & 99.76        & 88.68 & 94.79 & 98.95         \\
ResNet-20                & 79.6  & 99.22 & 99.77        & 79.66 & 91    & 98.34         \\
ResNet-32                & 86.83 & 99.14 & 99.51        & 65.51 & 97.47 & 98.35         \\
ResNet-44                & 65.24 & 99.03 & 99.54        & 77.92 & 87.34 & 99.09         \\
ResNet-56                & 85.7  & 99.04 & 99.76        & 74.33 & 95.48 & 98.98         \\
ShuffleNet V2 0.5$\times$     & 47.49 & 82.29 & 93.31        & 58.09 & 89.14 & 91.34         \\
ShuffleNet V2 1.0$\times$     & 66.08 & 86.98 & 95.9         & 74.26 & 90.98 & 90.64         \\
ShuffleNet V2 1.5$\times$     & 81.08 & 92.97 & 97.66        & 73.67 & 94.46 & 95.65         \\
ShuffleNet V2 2.0$\times$     & 66.88 & 90.17 & 97.29        & 80.18 & 95.36 & 94.16         \\
MobileNet V2 (0.5)      & 68.79 & 96.51 & 99.73        & 77.91 & 95.22 & 98.34         \\
MobileNet V2 (0.75)     & 71.79 & 97.9  & 99.04        & 77.78 & 97.24 & 93.57         \\
MobileNet V2 (1.0)      & 70.5  & 96.79 & 99.29        & 83.76 & 97.27 & 96.59         \\
MobileNet V2 (1.4)      & 74.65 & 96.75 & 99.12        & 91.22 & 99.11 & 98.78         \\
\bottomrule
\end{tabular}}}
\vspace{-4mm}
\end{table}
\section{Supplementary Evaluation Results}

\begin{table*}[!t]
\centering \footnotesize
\caption{Performance of multi-trigger CleanSheet.}
\label{TA:multi}
\vspace{2mm}
\begin{tabular}{ccccccccccc} 
\toprule
\multicolumn{11}{c}{CIFAR-10}                                                                         \\ 

class               & 0     & 1     & 2     & 3     & 4     & 5     & 6     & 7     & 8     & 9      \\ 
\midrule
VGG-11-BN           & 63.86 & 72.56 & 43.12 & 54.88 & 81.14 & 70.93 & 68.73 & 59.68 & 45.15 & 59.53  \\
VGG-13-BN           & 94.01 & 97.46 & 88.18 & 90.54 & 96.75 & 95.92 & 90.58 & 92.66 & 88.14 & 90.76  \\
VGG-16-BN           & 93.36 & 99.37 & 91.36 & 95.68 & 96.57 & 94.93 & 91.6  & 92.4  & 90.34 & 95.16  \\
VGG-19-BN           & 89.44 & 98.76 & 88.01 & 94.72 & 95.81 & 92.1  & 91.07 & 91.51 & 83.95 & 93.69  \\ 
RepVGG-A0        & 97.14 & 97.76 & 92.24 & 97.08 & 97.2  & 93.67 & 96.49 & 95.11 & 90.7  & 96.99  \\
RepVGG-A1          & 97.81 & 98    & 97.31 & 98.51 & 97.72 & 97.12 & 95.47 & 96.59 & 97.38 & 95.34  \\
RepVGG-A2          & 98.49 & 98.13 & 88.58 & 99.53 & 96.03 & 98.35 & 95.21 & 95.62 & 93.44 & 97.31  \\
ResNet-20            & 94.91 & 99.06 & 96.46 & 98.04 & 97.17 & 98.22 & 96.84 & 95.41 & 95.43 & 97.59  \\
ResNet-32            & 97.78 & 98.47 & 92.12 & 96.88 & 96.87 & 95.03 & 95.12 & 94.16 & 95.34 & 96.28  \\
ResNet-44            & 98.21 & 98.58 & 93.09 & 98.56 & 96.15 & 97.5  & 95.82 & 94.34 & 95.69 & 97.55  \\
ResNet-56            & 95.71 & 98.68 & 96.06 & 97.97 & 98.08 & 97.23 & 94.35 & 94.41 & 96.87 & 98.33  \\
ShuffleNet V2 0.5$\times$ & 84.13 & 93.41 & 61.62 & 54.91 & 71.62 & 60.6  & 86.84 & 74.05 & 75.48 & 65.71  \\
ShuffleNet V2 1.0$\times$ & 95.06 & 91.23 & 88.95 & 87.96 & 85.09 & 85.95 & 87.78 & 87.45 & 93.18 & 92.89  \\
ShuffleNet V2 1.5$\times$ & 95.07 & 95.42 & 65.77 & 90.28 & 76.62 & 92.66 & 92    & 87.13 & 86.4  & 89.39  \\
ShuffleNet V2 2.0$\times$ & 89.5  & 93.94 & 79.69 & 91.73 & 89.93 & 91.61 & 88.69 & 85.25 & 88.86 & 90.6   \\
MobileNet V2 (0.5)  & 97.06 & 96.62 & 88.87 & 91.6  & 97.3  & 89.6  & 96.66 & 94.88 & 96.76 & 97.26  \\
MobileNet V2 (0.75) & 99.17 & 96.63 & 91.16 & 96.04 & 96.3  & 96.51 & 97.36 & 90.18 & 93.96 & 97.34  \\
MobileNet V2 (1.0)  & 98.11 & 93.92 & 96.17 & 95.8  & 98.13 & 92.03 & 96.81 & 94.95 & 94.46 & 97.37  \\
MobileNet V2 (1.4)  & 98.44 & 97.55 & 92.75 & 97.26 & 95.15 & 94.86 & 96.5  & 95.94 & 96.77 & 97.54  \\

\toprule
\multicolumn{11}{c}{CIFAR-100}                                                                        \\ 
Class               & 1     & 11    & 21    & 31    & 41    & 51    & 61    & 71    & 81    & 91     \\ 
\midrule
VGG-11-BN           & 68.2  & 42.47 & 51.31 & 59.69 & 69.55 & 68.22 & 52.55 & 45.97 & 78.94 & 71.66  \\
VGG-13-BN           & 90.73 & 77.36 & 73.44 & 91.16 & 92.42 & 93.41 & 73.32 & 87.79 & 96.84 & 90.63  \\
VGG-16-BN           & 95.32 & 76.51 & 80.32 & 84.34 & 82.18 & 88.67 & 87.74 & 72.41 & 94.01 & 94.87  \\
VGG-19-BN           & 98.33 & 86.19 & 78.65 & 82.36 & 94.85 & 96.29 & 80.65 & 70.83 & 96.62 & 96.35  \\
RepVGG-A0        & 91.83 & 92.09 & 59.53 & 88.88 & 89.35 & 81.23 & 81.48 & 19.72 & 97.19 & 95.51  \\
RepVGG-A1          & 97.87 & 83.47 & 82.53 & 94.43 & 90.34 & 67.02 & 81.95 & 80.64 & 97.88 & 92.93  \\
RepVGG-A2          & 96.61 & 91.98 & 45.92 & 80.15 & 89.42 & 91.7  & 85.49 & 88.64 & 98.39 & 92.23  \\
ResNet-20            & 97.25 & 89.36 & 67.53 & 81.78 & 85.83 & 82.36 & 79.49 & 63.89 & 97.77 & 79.65  \\
ResNet-32            & 90.57 & 93.56 & 74.06 & 69.73 & 89.29 & 97.39 & 84.31 & 56.42 & 98.62 & 82.45  \\
ResNet-44            & 95.83 & 80.9  & 82.8  & 80.91 & 89.9  & 94.91 & 83.46 & 92.51 & 96.19 & 92.86  \\
ResNet-56            & 84.08 & 84.73 & 83.22 & 84.99 & 87.3  & 81.6  & 76.19 & 70.37 & 96.86 & 89.48  \\
ShuffleNet V2 0.5$\times$ & 82.66 & 56.31 & 36.86 & 69.35 & 45.16 & 70.18 & 64.83 & 23.08 & 81.15 & 75.12  \\
ShuffleNet V2 1.0$\times$ & 88.38 & 71.02 & 52.42 & 76.09 & 86.48 & 75.04 & 69.87 & 38.43 & 90.72 & 88.83  \\
ShuffleNet V2 1.5$\times$ & 92.73 & 77.38 & 71.44 & 74.15 & 71.22 & 80.24 & 78.29 & 52.48 & 98.06 & 91.91  \\
ShuffleNet V2 2.0$\times$ & 76.78 & 82.59 & 60.84 & 58.06 & 75.36 & 80.22 & 85.4  & 50.61 & 94.74 & 96.12  \\
MobileNet V2 (0.5)  & 90.47 & 86.06 & 40.64 & 74.9  & 87.32 & 88.65 & 74.21 & 77.14 & 95.08 & 93.3   \\
MobileNet V2 (0.75) & 84.25 & 75.82 & 63.86 & 86.72 & 88.01 & 87.66 & 81.96 & 93.94 & 96.22 & 89.93  \\
MobileNet V2 (1.0)  & 95.48 & 66.15 & 64.9  & 88.1  & 59.92 & 75.49 & 90.36 & 47.16 & 95.94 & 89.8   \\
MobileNet V2 (1.4)  & 91.47 & 94.02 & 75.73 & 83.75 & 86.31 & 92.94 & 88.41 & 82.5  & 96.09 & 96.1   \\
\bottomrule
\end{tabular}
\vspace{-4mm}
\end{table*}



\begin{table*}[t]
\centering \footnotesize
\caption{Performance of CleanSheet under different pruning ratios.}
\label{tab:prune}
\setlength{\tabcolsep}{3pt}{
\resizebox{\linewidth}{!}{
\begin{tabular}{ccccccccccccccc} 
\toprule
Pruning ratio       & \multicolumn{2}{c}{0} & \multicolumn{2}{c}{0.05} & \multicolumn{2}{c}{0.1} & \multicolumn{2}{c}{0.15} & \multicolumn{2}{c}{0.2} & \multicolumn{2}{c}{0.25} & \multicolumn{2}{c}{0.3}  \\
metrics             & CA(\%) & ASR(\%)      & CA(\%) & ASR(\%)         & CA(\%) & ASR(\%)        & CA(\%) & ASR(\%)         & CA(\%) & ASR(\%)        & CA(\%) & ASR(\%)         & CA(\%) & ASR(\%)         \\ 
\toprule

\multicolumn{15}{c}{CIFAR-10}                                                                                                         \\
MobileNet V2 (0.5)  & 93.12 & 97.60 & 91.57 & 96.99 & 69.74 & 0.02  & 63.10 & 0.00  & 61.12 & 0.00  & 37.89 & 0.00  & 23.78 & 0.00   \\
MobileNet V2 (0.75) & 94.08 & 97.86 & 93.64 & 96.88 & 82.92 & 1.27  & 72.60 & 19.63 & 67.27 & 8.54  & 61.51 & 2.02  & 40.31 & 0.00   \\
MobileNet V2 (1.0)  & 94.05 & 96.55 & 93.19 & 93.64 & 82.68 & 0.00  & 73.12 & 17.46 & 69.62 & 9.02  & 57.07 & 1.17  & 40.15 & 0.00   \\
MobileNet V2 (1.4)  & 99.83 & 98.65 & 94.22 & 98.47 & 84.33 & 1.34  & 78.68 & 56.17 & 73.77 & 55.98 & 71.75 & 23.63 & 56.12 & 10.22  \\
RepVGG-A0          & 94.47 & 99.18 & 94.43 & 99.15 & 88.72 & 50.30 & 82.77 & 73.95 & 82.14 & 68.34 & 80.80 & 66.28 & 69.38 & 53.52  \\
RepVGG-A1          & 94.93 & 98.48 & 94.94 & 98.46 & 90.52 & 62.96 & 82.06 & 0.00  & 81.87 & 0.00  & 81.69 & 0.00  & 72.08 & 0.00   \\
RepVGG-A2          & 95.27 & 98.59 & 95.27 & 98.60 & 89.05 & 60.87 & 77.93 & 58.58 & 76.81 & 44.39 & 75.86 & 30.23 & 72.92 & 77.81  \\
ResNet-20            & 92.59 & 99.12 & 48.93 & 96.18 & 12.69 & 0.00  & 10.00 & 0.00  & 10.00 & 0.00  & 10.00 & 0.00  & 10.00 & 0.00   \\
ResNet-32            & 93.53 & 98.91 & 80.74 & 94.09 & 32.80 & 0.00  & 42.13 & 58.55 & 20.13 & 0.00  & 10.84 & 0.00  & 10.74 & 0.00   \\
ResNet-44            & 94.01 & 98.72 & 85.97 & 90.48 & 58.19 & 0.00  & 31.03 & 0.00  & 23.63 & 0.00  & 11.15 & 0.00  & 11.48 & 0.00   \\
ResNet-56            & 94.38 & 99.10 & 87.45 & 96.93 & 58.34 & 0.00  & 21.70 & 0.00  & 13.21 & 0.00  & 10.38 & 0.00  & 10.01 & 0.00   \\
ShuffleNet V2 0.5$\times$ & 90.65 & 85.70 & 89.39 & 83.37 & 81.07 & 77.48 & 69.17 & 56.12 & 59.78 & 6.21  & 47.29 & 3.97  & 36.76 & 2.35   \\
ShuffleNet V2 1.0$\times$ & 93.31 & 88.61 & 92.75 & 86.68 & 83.87 & 1.81  & 74.43 & 1.27  & 70.28 & 0.15  & 56.42 & 0.01  & 33.94 & 0.00   \\
ShuffleNet V2 1.5$\times$ & 93.57 & 93.43 & 93.45 & 92.70 & 85.13 & 11.28 & 74.22 & 7.85  & 73.26 & 3.84  & 67.68 & 1.43  & 54.51 & 0.14   \\
ShuffleNet V2 2.0$\times$ & 93.98 & 89.55 & 93.94 & 87.49 & 87.62 & 22.03 & 80.79 & 32.05 & 79.00 & 23.10 & 74.18 & 10.86 & 63.20 & 5.51   \\
VGG-11-BN           & 92.78 & 71.16 & 90.52 & 63.07 & 78.33 & 0.28  & 61.64 & 0.05  & 49.04 & 0.00  & 41.03 & 0.00  & 35.69 & 0.00   \\
VGG-13-BN           & 94.00 & 98.35 & 92.20 & 97.99 & 74.24 & 94.60 & 54.97 & 0.20  & 33.93 & 0.00  & 13.74 & 0.00  & 10.20 & 0.00   \\
VGG-16-BN           & 94.15 & 99.04 & 92.77 & 97.42 & 80.90 & 97.35 & 60.40 & 97.08 & 36.01 & 76.55 & 18.57 & 0.31  & 12.64 & 0.00   \\
VGG-19-BN           & 93.91 & 98.98 & 92.59 & 97.94 & 84.15 & 98.32 & 60.58 & 96.94 & 43.91 & 91.04 & 23.47 & 35.73 & 10.00 & 0.00   \\ 
\midrule
\multicolumn{15}{c}{CIFAR-100}                                                                                                        \\ 

MobileNet V2 (0.5)  & 71.15 & 98.37 & 61.02 & 98.66 & 44.21 & 98.69 & 21.93 & 99.89 & 6.25  & 58.99 & 2.62  & 23.41 & 1.46  & 62.08  \\
MobileNet V2 (0.75) & 74.10 & 94.11 & 66.63 & 97.97 & 55.07 & 96.81 & 32.64 & 82.82 & 11.04 & 59.34 & 5.08  & 57.11 & 1.97  & 0.03   \\
MobileNet V2 (1.0)  & 74.30 & 97.43 & 67.93 & 98.13 & 57.97 & 97.38 & 42.88 & 95.74 & 32.11 & 90.21 & 11.62 & 36.17 & 4.95  & 0.02   \\
MobileNet V2 (1.4)  & 76.33 & 99.19 & 70.14 & 99.39 & 63.75 & 99.39 & 52.65 & 99.12 & 41.71 & 99.41 & 24.42 & 97.93 & 8.98  & 0.00   \\
RepVGG-A0          & 75.29 & 98.23 & 70.34 & 99.00 & 64.08 & 98.73 & 55.97 & 97.89 & 48.44 & 97.74 & 38.02 & 96.30 & 27.43 & 0.00   \\
RepVGG-A1          & 76.45 & 99.38 & 71.11 & 99.12 & 66.49 & 0.00  & 60.06 & 0.00  & 53.68 & 0.00  & 27.70 & 0.00  & 18.39 & 0.00   \\
RepVGG-A2          & 77.49 & 98.68 & 72.79 & 98.73 & 68.00 & 98.86 & 62.89 & 98.63 & 57.04 & 98.08 & 48.30 & 0.00  & 29.71 & 0.00   \\
ResNet-20            & 68.84 & 99.09 & 16.74 & 55.83 & 2.35  & 0.00  & 1.71  & 0.00  & 1.72  & 0.00  & 1.10  & 0.00  & 1.83  & 0.00   \\
ResNet-32            & 70.14 & 96.86 & 18.47 & 45.29 & 1.40  & 0.00  & 1.73  & 0.00  & 1.96  & 0.00  & 1.27  & 0.00  & 1.00  & 0.00   \\
ResNet-44            & 71.65 & 98.64 & 1.28  & 0.00  & 1.37  & 0.00  & 1.00  & 0.00  & 1.00  & 0.00  & 1.03  & 0.00  & 0.99  & 0.00   \\
ResNet-56            & 72.61 & 96.67 & 10.90 & 0.00  & 2.33  & 0.00  & 1.76  & 0.00  & 2.10  & 0.00  & 1.27  & 0.00  & 1.00  & 0.00   \\
ShuffleNet V2 0.5$\times$ & 67.82 & 88.40 & 57.46 & 90.07 & 41.60 & 74.38 & 21.68 & 49.86 & 6.15  & 48.84 & 2.40  & 0.00  & 1.00  & 0.00   \\
ShuffleNet V2 1.0$\times$ & 72.64 & 94.55 & 66.48 & 90.82 & 59.05 & 92.51 & 49.35 & 57.87 & 37.05 & 52.64 & 23.23 & 53.73 & 8.74  & 0.00   \\
ShuffleNet V2 1.5$\times$ & 74.23 & 96.79 & 69.20 & 97.73 & 63.13 & 95.98 & 56.25 & 93.42 & 45.29 & 89.13 & 33.28 & 0.00  & 19.28 & 0.00   \\
ShuffleNet V2 2.0$\times$ & 75.49 & 91.46 & 70.56 & 89.21 & 65.52 & 93.59 & 59.22 & 95.06 & 51.86 & 87.26 & 38.48 & 0.00  & 24.13 & 0.00   \\
VGG-11-BN           & 70.79 & 83.71 & 61.81 & 83.04 & 50.64 & 85.16 & 27.79 & 0.00  & 15.57 & 0.00  & 4.82  & 0.00  & 2.86  & 0.00   \\
VGG-13-BN           & 74.63 & 97.12 & 62.03 & 98.36 & 44.91 & 0.00  & 20.76 & 0.00  & 5.80  & 0.00  & 2.03  & 0.00  & 1.27  & 0.00   \\
VGG-16-BN           & 74.00 & 97.98 & 62.73 & 98.20 & 24.16 & 95.92 & 2.77  & 0.00  & 1.67  & 0.00  & 1.00  & 0.00  & 1.12  & 0.00   \\
VGG-19-BN           & 73.83 & 99.45 & 64.91 & 99.45 & 39.57 & 99.05 & 8.66  & 94.45 & 1.58  & 0.00  & 1.01  & 0.00  & 1.00  & 0.00   \\
\midrule

\multicolumn{15}{c}{GTSRB}                                                                                                           \\
ResNet-18            & 96.44 & 92.80 & 96.44 & 92.91 & 91.55 & 93.59 & 87.33 & 93.98 & 87.31 & 94.50 & 87.58 & 95.14 & 81.40 & 95.93  \\
ResNet-34            & 96.37 & 96.52 & 96.38 & 96.61 & 95.15 & 96.66 & 89.35 & 96.99 & 89.38 & 97.56 & 89.33 & 97.78 & 83.06 & 98.35  \\
ResNet-50            & 96.44 & 92.80 & 96.44 & 92.91 & 91.55 & 93.59 & 87.33 & 93.98 & 87.31 & 94.50 & 87.58 & 95.14 & 81.40 & 95.93  \\
RepVGG-A0          & 96.55 & 97.07 & 96.55 & 97.07 & 90.73 & 97.46 & 87.84 & 97.56 & 87.88 & 97.59 & 87.82 & 97.63 & 81.95 & 97.80  \\
RepVGG-A1          & 96.49 & 96.79 & 96.49 & 96.80 & 93.87 & 96.97 & 85.86 & 97.20 & 86.06 & 97.24 & 86.25 & 97.26 & 82.24 & 97.37  \\
RepVGG-A2          & 96.65 & 96.46 & 96.65 & 96.46 & 91.81 & 96.82 & 88.25 & 96.91 & 88.32 & 96.93 & 88.36 & 96.95 & 82.24 & 97.20  \\
MobileNet V2 (0.5)  & 92.69 & 91.04 & 92.62 & 90.71 & 87.40 & 91.61 & 82.10 & 92.45 & 81.22 & 92.89 & 80.22 & 93.93 & 72.39 & 94.08  \\
MobileNet V2 (0.75) & 95.58 & 95.37 & 95.61 & 95.52 & 89.78 & 95.24 & 84.49 & 95.62 & 83.97 & 95.39 & 83.36 & 96.74 & 74.72 & 97.76  \\
MobileNet V2 (1.4)  & 95.61 & 95.69 & 95.58 & 95.64 & 89.92 & 95.54 & 85.59 & 95.53 & 85.36 & 95.56 & 85.36 & 95.57 & 79.53 & 95.72  \\
MobileNet V2 (1.0)  & 95.53 & 96.50 & 95.55 & 96.34 & 89.94 & 96.18 & 84.65 & 95.84 & 84.56 & 95.37 & 84.29 & 93.55 & 78.10 & 93.13  \\
ShuffleNet V2 0.5$\times$ & 95.19 & 92.63 & 95.21 & 92.47 & 89.83 & 92.52 & 83.83 & 93.75 & 83.82 & 93.04 & 83.39 & 91.16 & 75.86 & 78.06  \\
ShuffleNet V2 1.5$\times$ & 95.23 & 96.01 & 95.24 & 96.05 & 90.75 & 96.35 & 85.56 & 96.50 & 85.67 & 96.62 & 85.79 & 96.45 & 79.57 & 96.96  \\
ShuffleNet V2 1.0$\times$ & 95.83 & 93.64 & 95.83 & 93.65 & 90.66 & 93.83 & 85.23 & 94.08 & 85.14 & 93.69 & 84.78 & 94.25 & 78.47 & 94.65  \\
ShuffleNet V2 2.0$\times$ & 95.75 & 93.10 & 95.75 & 93.10 & 90.72 & 93.18 & 84.70 & 93.48 & 84.86 & 93.92 & 84.79 & 94.38 & 78.38 & 95.00  \\

\midrule

\multicolumn{15}{c}{SVHN}                                                                                                           \\   
MobileNet V2 (0.5)  & 97.26 & 93.92 & 96.29 & 93.66 & 95.38 & 93.64 & 93.63 & 93.11 & 91.35 & 94.39 & 89.67 & 94.19 & 85.98 & 91.11  \\
MobileNet V2 (0.75) & 98.14 & 93.63 & 96.99 & 93.27 & 95.99 & 92.61 & 94.52 & 91.53 & 92.11 & 90.63 & 90.55 & 89.75 & 87.61 & 91.61  \\
MobileNet V2 (1.0)  & 97.93 & 90.46 & 97.15 & 90.01 & 96.01 & 90.10 & 94.61 & 88.32 & 93.03 & 86.32 & 91.20 & 87.95 & 88.86 & 86.46  \\
MobileNet V2 (1.4)  & 97.77 & 98.03 & 96.71 & 98.15 & 95.47 & 98.31 & 94.74 & 97.84 & 92.31 & 97.27 & 90.29 & 97.10 & 87.65 & 96.79  \\
RepVGG-A0          & 98.15 & 92.48 & 97.21 & 93.22 & 96.33 & 93.84 & 95.46 & 94.28 & 93.34 & 94.22 & 92.29 & 94.39 & 90.55 & 95.04  \\
RepVGG-A1          & 98.00 & 94.39 & 97.15 & 94.78 & 95.87 & 95.32 & 95.09 & 95.34 & 93.07 & 95.77 & 91.77 & 96.45 & 90.07 & 96.75  \\
RepVGG-A2          & 98.43 & 93.37 & 97.47 & 93.88 & 96.30 & 94.73 & 95.04 & 95.64 & 93.63 & 96.75 & 92.38 & 97.87 & 90.72 & 98.38  \\
ResNet-18            & 98.23 & 91.61 & 97.25 & 92.16 & 95.97 & 92.98 & 94.73 & 93.63 & 93.48 & 94.14 & 91.84 & 93.63 & 89.88 & 95.25  \\
ResNet-34            & 97.51 & 90.70 & 96.59 & 90.51 & 95.31 & 89.77 & 93.89 & 89.99 & 92.34 & 89.07 & 91.19 & 91.28 & 88.99 & 90.48  \\
ResNet-50            & 97.62 & 91.31 & 96.70 & 91.87 & 95.53 & 92.45 & 94.25 & 93.43 & 92.78 & 94.08 & 91.12 & 95.46 & 89.13 & 95.52  \\
ShuffleNet V2 0.5$\times$ & 97.43 & 85.32 & 96.51 & 85.58 & 95.46 & 86.61 & 93.87 & 86.75 & 91.95 & 87.37 & 90.19 & 85.45 & 87.36 & 85.97  \\
ShuffleNet V2 1.0$\times$ & 97.78 & 90.46 & 96.79 & 90.32 & 95.57 & 90.56 & 94.38 & 90.40 & 92.57 & 91.00 & 91.20 & 90.81 & 89.62 & 90.53  \\
ShuffleNet V2 1.5$\times$ & 97.68 & 93.44 & 96.77 & 93.43 & 95.88 & 93.75 & 94.57 & 93.48 & 93.26 & 93.71 & 91.98 & 93.92 & 90.09 & 93.56  \\
ShuffleNet V2 2.0$\times$ & 97.91 & 85.55 & 96.98 & 85.64 & 95.84 & 85.90 & 94.69 & 85.95 & 92.75 & 86.25 & 91.60 & 89.29 & 90.22 & 90.69 \\

\bottomrule
\end{tabular}}}
\vspace{-4mm}
\end{table*}

\subsection{Multi-trigger Attacks}\label{se:MT}
In Table~\ref{TA:multi}, we provide more results for the multi-trigger variations of CleanSheet against models with different architectures. For CIFAR-10, triggers were generated for five selected target classes. Then, we use the five class-related triggers to attack various clean models. The average ASR for the five classes is higher than 90\%, showing the strong performance of CleanSheet. This approach allows the adversary to manipulate the model to generate any desired output by introducing various triggers into different instances. Besides, the multi-trigger attacks do not require the adversaries to add their budget, as the multiple class-related triggers are embedded into the clean model during the training process.

\subsection{More Results about Model Pruning}\label{se:pr}
Table~\ref{tab:prune} presents the CAs and ASRs of the target models under different pruning rates on CIFAR-10. As shown, CleanSheet is robust against model pruning. In most cases, the ASRs are unaffected when some weights are removed. These results show that CleanSheet leverages crucial weights to model classification, as the key features in our trigger are strongly related to specific classes.
In certain cases, such as with ShuffleNet V2 0.5$\times$, both CA and ASR decrease simultaneously. We attribute this to the pruning of certain key weights associated with classification.

\begin{figure*}[!t]
    \centering
    \includegraphics[width=0.8\linewidth]{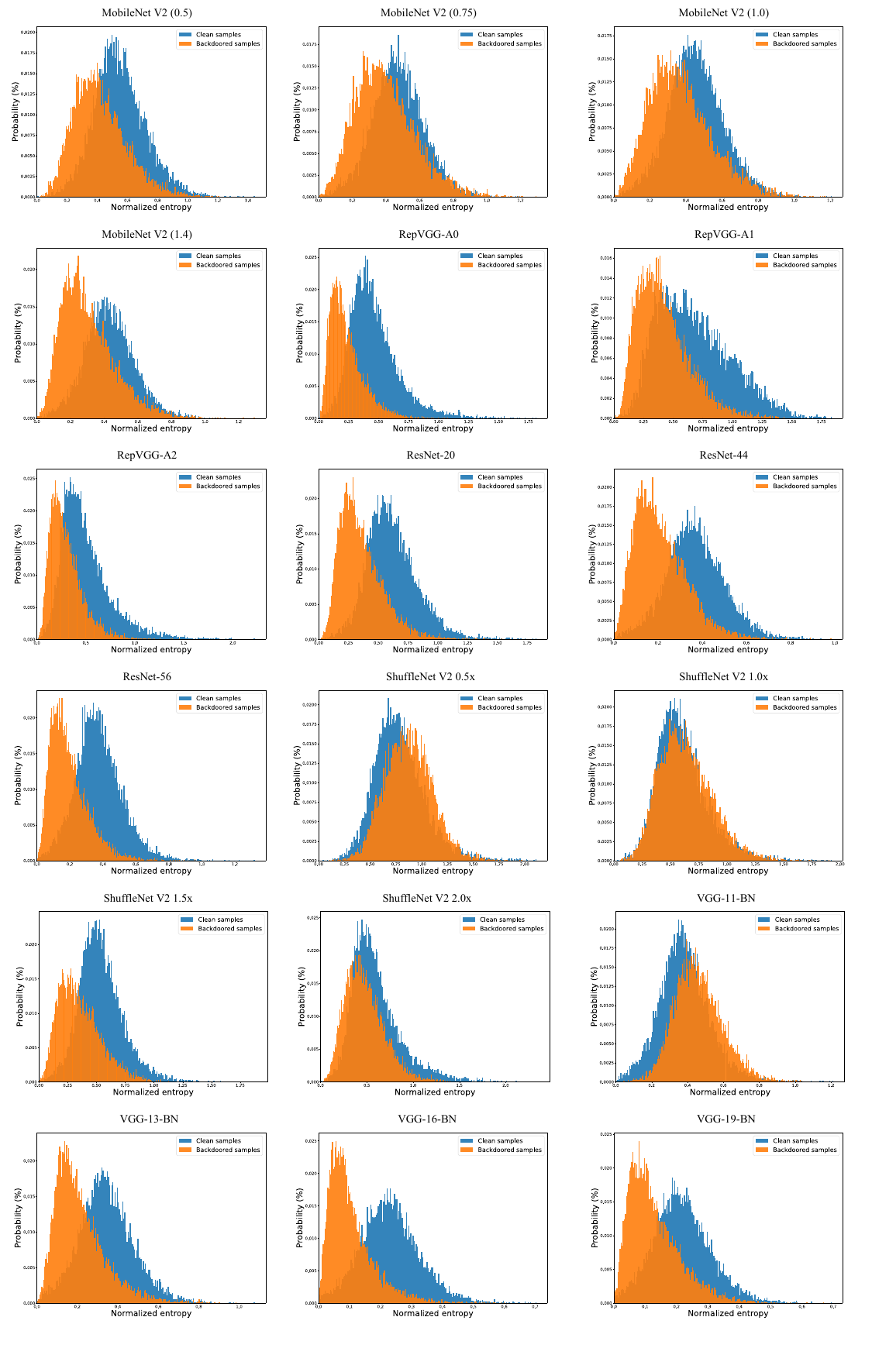}
    \caption{The distribution of regularized entropy for clean samples and backdoored samples of CIFAR-10 under Strip detection scheme.}
    \label{fig:strip}
\end{figure*}

\end{document}